%% file: main.tex
\documentclass[conference]{IEEEtran}
\IEEEoverridecommandlockouts
\input{tool}

\begin{document}

\title{\model: Reinforcement Learning for Repository-Level Code Completion}
% \author{\IEEEauthorblockN{Anonymous author}}

\author{
    \IEEEauthorblockN{Yanlin Wang$^{1}$, Yanli Wang$^{1}$, Daya Guo$^{1}$, Jiachi Chen$^{1*}$\thanks{* Corresponding author}, Ruikai Zhang$^{2}$, Yuchi Ma$^{2}$, Zibin Zheng$^{1}$}
    \IEEEauthorblockA{$^{1}$ Sun Yat-sen University, Zhuhai, China}
    \IEEEauthorblockA{
    \{\href{mailto:wangylin36@mail.sysu.edu.cn}{wangylin36}, \href{mailto:chenjch86@mail.sysu.edu.cn}{chenjch86},
    \href{mailto:zhzibin@mail.sysu.edu.cn}{zhzibin}\}@mail.sysu.edu.cn,
    \{\href{mailto:wangyli58@mail2.sysu.edu.cn}{wangyli58}, \href{mailto:guody5@mail2.sysu.edu.cn}{guody5}\}@mail2.sysu.edu.cn,
    % \{\href{mailto:zhangruikai1@huawei.com}{zhangruikai1}, \href{mailto:mayuchi1@huawei.com}{mayuchi1}\}@huawei.com
    }
    \IEEEauthorblockA{$^{2}$ Huawei Cloud Computing Technologies Co., Ltd., Shenzhen, China}
    \IEEEauthorblockA{
    % \{\href{mailto:wangylin36@mail.sysu.edu.cn}{wangylin36}, \href{mailto:chenjch86@mail.sysu.edu.cn}{chenjch86},
    % \href{mailto:zhzibin@mail.sysu.edu.cn}{zhzibin}\}@mail.sysu.edu.cn,
    % \{\href{mailto:wangyli58@mail2.sysu.edu.cn}{wangyli58}, \href{mailto:guody5@mail2.sysu.edu.cn}{guody5}\}@mail2.sysu.edu.cn,
    \{\href{mailto:zhangruikai1@huawei.com}{zhangruikai1}, \href{mailto:mayuchi1@huawei.com}{mayuchi1}\}@huawei.com
    }
}

% \author{
% \IEEEauthorblockN{Yanlin Wang}
% \IEEEauthorblockA{\textit{School of Software Engineering} \\
% \textit{Sun Yat-sen University}\\
% Zhuhai, China \\
% yanlin-wang@outlook.com}
% \and
% \IEEEauthorblockN{Yanli Wang}
% \IEEEauthorblockA{\textit{School of Software Engineering} \\
% \textit{Sun Yat-sen University}\\
% Zhuhai, China \\
% wangyli58@mail2.sysu.edu.cn}
% \and
% \IEEEauthorblockN{Daya Guo}
% \IEEEauthorblockA{\textit{School of Computer Science} \\
% \textit{Sun Yat-sen University}\\
% Guangzhou, China \\
% guody5@mail2.sysu.edu.cn}
% \and
% \IEEEauthorblockN{Jiachi Chen*\thanks{* Corresponding author}}
% \IEEEauthorblockA{\textit{School of Software Engineering} \\
% \textit{Sun Yat-sen University}\\
% Zhuhai, China \\
% chenjch86@mail.sysu.edu.cn}
% \and
% \IEEEauthorblockN{Ruikai Zhang}
% \IEEEauthorblockA{\textit{Huawei Cloud Computing} \\
% \textit{Technologies Co., Ltd.}\\
% Shenzhen, China \\
% zhangruikai1@huawei.com}
% \and
% \IEEEauthorblockN{Yuchi Ma}
% \IEEEauthorblockA{\textit{Huawei Cloud Computing} \\
% \textit{Technologies Co., Ltd.}\\
% Shenzhen, China \\
% mayuchi1@huawei.com}
% \and
% \IEEEauthorblockN{Zibin Zheng}
% \IEEEauthorblockA{\textit{School of Software Engineering} \\
% \textit{Sun Yat-sen University}\\
% Zhuhai, China \\
% zhzibin@mail.sysu.edu.cn}
% }

\maketitle

\begin{abstract}

Repository-level code completion aims to generate code for unfinished code snippets within the context of a specified repository. Existing approaches mainly rely on retrieval-augmented generation strategies due to limitations in input sequence length. However, traditional lexical-based retrieval methods like BM25 struggle to capture code semantics, while model-based retrieval methods face challenges due to the lack of labeled data for training.
Therefore, we propose RLCoder, a novel reinforcement learning framework, which can enable the retriever to learn to retrieve useful content for code completion without the need for labeled data.
Specifically, we iteratively evaluate the usefulness of retrieved content based on the perplexity of the target code when provided with the retrieved content as additional context, and provide feedback to update the retriever parameters.  This iterative process enables the retriever to learn from its successes and failures, gradually improving its ability to retrieve relevant and high-quality content.
Considering that not all situations require information beyond code files and not all retrieved context is helpful for generation, we also introduce a stop signal mechanism, allowing the retriever to decide when to retrieve and which candidates to retain autonomously.
Extensive experimental results demonstrate that RLCoder consistently outperforms state-of-the-art methods on CrossCodeEval and RepoEval, achieving 12.2\% EM improvement over previous methods. 
Moreover, experiments show that our framework can generalize across different programming languages and further improve previous methods like RepoCoder.
\end{abstract}


\input{body}

% \newpage
\bibliographystyle{IEEEtran}
\bibliography{ref}

\end{document}

%% file: tool.tex
\usepackage{cite}
\usepackage{amsmath,amssymb,amsfonts}
\usepackage{algorithmic}
\usepackage{graphicx}
\usepackage{textcomp}
\usepackage{colortbl}  %彩色表格需要加载的宏包
\usepackage{xcolor}
\usepackage{multirow} % 引入multirow包
\usepackage{booktabs} % 提供\toprule等命令
\usepackage{hyperref}
\usepackage{algorithm}
\usepackage{tcolorbox}
\usepackage{tikz}
\usepackage{threeparttable}

\usepackage{enumitem}
\usepackage{fmtcount}

\usepackage{latexsym}
\usepackage{url}
\usepackage{arydshln}

\def\BibTeX{{\rm B\kern-.05em{\sc i\kern-.025em b}\kern-.08em
    T\kern-.1667em\lower.7ex\hbox{E}\kern-.125emX}}

\usepackage{xspace}

\newcommand{\mytextsuperscript}[1]{\textsuperscript{#1}}

\newcommand{\model}{{RLCoder}\xspace}
\newcommand{\rlretriever}{RLRetriever\xspace}

\newcommand{\RawragCLSeven}{RawRAG\textsubscript{ CodeLlama-7B}}
\newcommand{\RepocoderCLSeven}{RepoCoder\textsubscript{ CodeLlama-7B}}
\newcommand{\RlcoderCLSeven}{RLCoder\textsubscript{ CodeLlama-7B}}

\newcommand{\RawragSCSeven}{RawRAG\textsubscript{ StarCoder-7B}}
\newcommand{\RepocoderSCSeven}{RepoCoder\textsubscript{ StarCoder-7B}}
\newcommand{\RlcoderSCSeven}{RLCoder\textsubscript{ StarCoder-7B}}

\newcommand{\RawragSCTSeven}{RawRAG\textsubscript{ StarCoder2-7B}}
\newcommand{\RepocoderSCTSeven}{RepoCoder\textsubscript{ StarCoder2-7B}}
\newcommand{\RlcoderSCTSeven}{RLCoder\textsubscript{ StarCoder2-7B}}

\newcommand{\RawragDSCOne}{RawRAG\textsubscript{ DeepSeekCoder-1B}}
\newcommand{\RepocoderDSCOne}{RepoCoder\textsubscript{ DeepSeekCoder-1B}}
\newcommand{\RlcoderDSCOne}{RLCoder\textsubscript{ DeepSeekCoder-1B}}

\newcommand{\RawragDSCSeven}{RawRAG\textsubscript{ DeepSeekCoder-7B}}
\newcommand{\RepocoderDSCSeven}{RepoCoder\textsubscript{ DeepSeekCoder-7B}}
\newcommand{\RlcoderDSCSeven}{RLCoder\textsubscript{ DeepSeekCoder-7B}}

\newcommand{\Baseline}{NoRetrieval} %Baseline
\newcommand{\BM}{BM25}
\newcommand{\UniXcoder}{UniXcoder}
\newcommand{\UniXcoderSFT}{UniXcoder-SFT}
\newcommand{\RLRetriever}{RLRetreiver}

\newcommand{\boxmargin}{1mm}
\tcbuselibrary{most, skins, breakable, theorems}
\newtcolorbox{myboxa}[2][]{
    colback=gray!10!white,
    colframe=black, enhanced,
    attach boxed title to top left={yshift=-2mm,xshift=5mm},
    title=#2,#1
}
\newtcolorbox{myboxb}[2][]{
    boxsep=3pt,
    left = \boxmargin, right = \boxmargin, top = \boxmargin, bottom = \boxmargin,
    title={#2},#1
}
\newtcolorbox{myboxc}{
    colback=gray!15!white,
    arc = 0pt, outer arc = 0pt,
    boxsep=0pt, left = 3pt, right = 0pt, top = 0pt, bottom = 0pt, 
    leftrule=3pt, bottomrule=0pt,toprule=0pt, rightrule=0pt,
    left = \boxmargin, right = \boxmargin, top = \boxmargin, bottom = \boxmargin
}
\newtcolorbox{myboxd}{
    colback=gray!10,%gray background
    colframe=black,% black frame colour
    width=\columnwidth,% Use 8cm total width,
    arc=1mm, auto outer arc,
    boxrule=0.5pt,
}

%% file: body.tex
\section{Introduction}
With the advancement of large language models for code (code LLMs)~\cite{chen2021evaluating,nijkamp2023codegen,rozière2024code,li2023starcoder,guo2024deepseekcoder}, code completion has emerged as one of the most important features in integrated development environments (IDEs)~\cite{liu2024non,wang2023practitioners,hellendoorn2019code,izadi2022codefill,zhou2022improving,izadi2024language}. However, due to the vast size of code repositories and the limitations of context length in models, repository-level code completion, which involves generating code suggestions within the context of an entire repository, cannot practically leverage the entire repository directly as context~\cite{ding2023cocomic}.
Therefore, previous works~\cite{ding2023cocomic,zhang2023repocoder,bairi2023codeplan,liang2024repofuse,eghbali2024dehallucinator,zhang2024codeagent} typically employ a retrieval-augmented-generation (RAG) strategy. In this approach, the unfinished code in the current file serves as a query to retrieve code candidates from the entire repository, providing cross-file context. These candidates are then concatenated with the unfinished code before being fed into code LLMs. %To retrieve relevant code snippets from other files, CodePlan~\cite{bairi2023codeplan} and RepoFuse~\cite{liang2024repofuse} employ static code analysis. RepoCoder~\cite{zhang2023repocoder} and De-Hallucinator~\cite{eghbali2024dehallucinator} adopt an approach through iterative retrieval and generation. CodeAgent~\cite{zhang2024codeagent} and ToolGen~\cite{wang2024teaching} explore tool invocation to help code generation. CoCoMIC~\cite{ding2023cocomic} and RepoHyper~\cite{phan2024repohyper} combine dependency analysis and learning methods to parse cross-file entities. 
To retrieve relevant code snippets from other files, various retrievers are adopted. RepoFuse~\cite{liang2024repofuse} uses lexical-based method BM25~\cite{robertson2009BM25} as the retriever to retrieve code snippets that are textually similar with the unfinished code. RepoCoder~\cite{zhang2023repocoder} and RepoHyper~\cite{phan2024repohyper} use the model-based approach that encodes code candidates and unfinished code into vectors and employs dense retrieval to find similar codes. Although these efforts have shown promising performance in repository-level code generation, we have identified the following problems in retrieval.

\begin{enumerate}[label={\bfseries P\arabic*}]
\item \textbf{Labeled Data Dependency.} Lexical-based methods such as BM25~\cite{robertson2009BM25} cannot capture code semantics, while model-based methods~\cite{ding2023cocomic,phan2024repohyper,wu2024repoformer} are capable of understanding code semantics but are hampered by the lack of ground-truth candidate data for training. This labeled data is hard to obtain, as it requires significant effort in data parsing and expert labeling, limiting its generalizability. 
\item \textbf{Candidate Construction Issue.} Previous methods of code candidate construction mainly employ the fixed window strategy~\cite{zhang2023repocoder} or dependency parsing~\cite{ding2023cocomic,liao2024a3codgen}. However, the fixed window strategy may disrupt the continuity of the code. Methods based on dependency parsing can only focus on limited context in the dependency graph and can not be applied to complex scenarios.
% \item \textbf{Non-selective retrieval.} Previous studies typically directly retrieve several candidates to serve as a context for a generation. However, the retrieved contents are not always necessary, and they can even detract from the performance in completion scenarios that do not require repository context. 
\item \textbf{Non-Selective Retrieval.} Previous studies typically directly retrieve several candidates to serve as the context for generation, neglecting when to retrieve and which candidates to retain. Unnecessary candidates can detract from the performance in completion scenarios that do not require repository context.
\end{enumerate}
%(1) The lack of labeled data limits previous learning-based methods. These methods usually require significant human effort in data parsing and are difficult to extend to other languages or repositories. (2) Previous studies on candidate construction primarily employ fixed windows and dependency parsing. However, the fixed window method may disrupt the continuity of the code, and dependency parsing can only focus on limited context in the dependency graph. (3) Previous studies directly use the top \(k\) retrieved candidates as context for generation. However, not all candidates contribute to the generation.

%In this paper, we propose a reinforcement learning framework for repository-level code completion named \textsc{RLCoder}. Unlike supervised learning methods, RLCoder learns by obtaining feedback from the generator without needing labeled data. Specifically, we use the retrieved candidates as context for the generator and calculate perplexity (PPL) to generate the unfinished target code. Considering that hallucinations in repository-level code completion often result from the use of incorrect identifiers or APIs, we assign higher weights to these specific tokens' PPL. The weighted PPL is then used as reward to provide feedback to the retriever. To better adapt to general scenarios, we introduce a stop signal in the training stage to avoid unnecessary cross-file context, thereby utilizing more in-file information. Besides, we introduce a simple yet effective Split-Aggregate candidate construction method based on human programming habits. 

In this paper, we propose \textsc{RLCoder}, a reinforcement learning framework for repository-level code completion to address the aforementioned problems. 
Firstly, we propose a codebase construction pipeline with a simple yet effective Split-Aggregate strategy. This approach allows better code continuity of the candidates, which we refer to as natural candidates (addressing \textbf{P2}). 
Secondly, during the training stage, we diverge from supervised learning methods that depend on labeled data. Instead, we train a retriever named RLRetriever that learns what to retrieve based on feedback from a specifically designed evaluator, without needing labeled data (addressing \textbf{P1}). Specifically, we iteratively evaluate the usefulness of retrieved content based on the perplexity of the target code when provided with the retrieved content as additional context, and provide feedback to update the retriever parameters, which enables the retriever to learn from its successes and failures, gradually improving its ability to retrieve relevant and high-quality content. Moreover, to mitigate hallucinations often observed in repository-level code completions, typically due to incorrect identifier or API usage~\cite{eghbali2024dehallucinator}, we design a weighted perplexity (PPL) mechanism that allocates higher weights to certain important tokens in perplexity calculation. Furthermore, considering that not all candidates retrieved are useful for generation, we introduce a stop signal mechanism to evaluate the usefulness of candidates, allowing the retriever to decide when to retrieve and which candidates to retain autonomously (addressing \textbf{P3}).
% additional information Those do not contribute positively to the generation will be removed, ensuring that only helpful context is retained for more efficient and effective code completion  
% Considering that not all situations require external code file information, we introduce a stop signal mechanism to avoid unnecessary cross-file context, which can allow the retriever to autonomously decide whether to retrieve additional information (addressing \textbf{P3}). 
Finally, in the inference stage, given an unfinished code as input, RLCoder retrieves natural candidates from the codebase, retains the useful candidates, and then feeds them along with the unfinished code into the generator (a backbone LLM) for target code generation.

% Firstly, unlike supervised learning methods which depend on labeled data, RLCoder learns what to retrieve by obtaining feedback from a designed evaluator, without needing labeled data. Specifically, we calculate the perplexity of the target code, given the retrieved candidates concatenated with unfinished code. The perplexity of the target code serves as a reward, providing feedback to train the retriever. Moreover, considering that hallucinations in repository-level code completion often result from the use of incorrect identifiers or APIs~\cite{eghbali2024dehallucinator}, we employ a weighted mechanism that allocates higher weights to certain important tokens when calculating the perplexity of the target code.
% Secondly, we propose a simple yet effective Split-Aggregate candidate construction method with better code continuity. 
% Thirdly, considering that not all situations require information beyond code files, we introduce a stop signal mechanism to avoid unnecessary cross-file context, which can allow the retriever to autonomously decide whether to retrieve additional information.

% We evaluate \model with extensive experiments with several LLMs on various benchmarks including CrossCodeEval~\cite{ding2023crosscodeeval}, RepoEval~\cite{zhang2023repocoder}, %CoderEval~\cite{Yu2023CoderEvalAB}, 
% and a new benchmark. 
We evaluate \model with extensive experiments with several LLMs on  CrossCodeEval~\cite{ding2023crosscodeeval} and RepoEval~\cite{zhang2023repocoder}. 
Experimental results show that our framework achieves 12.2\% improvement of Exact Match compared with previous methods. Furthermore, \model demonstrates high generalizability, showing effectiveness across various LLMs and programming languages. Additionally, experiments show that \model %is scalable, it 
can be integrated into previous methods such as RepoCoder to enhance code completion performance further.

Our main contributions are:
\begin{itemize}
    \item We propose \model, a reinforcement learning framework for repository-level code completion. To our knowledge, we are the first to train the retriever without labeled data for repository-level code completion. Besides, we design a mechanism that uses the weighted perplexity of the target code as the reward to further enhance performance.
    \item We introduce a simple yet effective Split-Aggregate candidate construction strategy based on human programming habits. This method avoids the disruption of code continuity and outperforms fixed window candidates indicated by the experimental results.
    \item We propose a stop signal mechanism to evaluate the usefulness of candidates and discard useless candidates for more effective code completion.
    \item We perform an extensive evaluation of RLCoder. Experimental results show that RLCoder outperforms the state-of-the-art methods and demonstrates generalizability and applicability. We provide the code and data at  \url{https://github.com/DeepSoftwareAnalytics/RLCoder}.
\end{itemize}

\section{Background}

\subsection{Retrieval-Augmented Generation}
Retrieval-augmented generation (RAG)~\cite{gao2024retrievalaugmented} is an approach that enhances the quality of generation by retrieving from external knowledge bases. This method includes three key components~\cite{zhao2024retrievalaugmented}: \textit{retriever}, \textit{generator}, and \textit{augmentation techniques}. The \textit{retriever} is used to find relevant information from a large-scale dataset or knowledge base, including pertinent documents, facts, or text snippets that are relevant to the input query or prompt. The retrieved information is fed into the \textit{generator}, which integrates this external knowledge into the generation stage. \textit{Augmentation techniques} focus on how retrieved information is integrated into the generation process. To formalize the RAG process, consider a scenario where we want to generate code based on a query \(q\) and a set of retrieved candidates \(\{c_1, c_2, ..., c_n\}\). The process can be described by the following formula:
\[
\text{Code} = \text{Generate}\left( q, \text{Retrieve}(q, \{c_1, c_2, ..., c_n\}) \right) \tag{1}
\]
where the \(\text{Retrieve}(\cdot)\) function selects the most relevant candidates based on the query \(q\) from the candidate set \(\{c_1, c_2, ..., c_n\}\), and the \(\text{Generate}(\cdot)\) function then takes the query and the retrieved candidates to generate the target code.

In recent years, researchers have conducted a substantial amount of research related to RAG, highlighting its promising potential for future applications~\cite{gao2024retrievalaugmented,cao2024retrieval,he2023rest,Nashid2023RetrievalBasedPS,liu2022uniparser}. Many studies have utilized RAG for code-related research~\cite{lu2022reacc,yu2022bashexplainer,Li_2021,zhang2023syntax,Zhang2020neuralsourcecodesummarization,li2023acecoder,parvez2021rag,zhou2023docpromptingrag,zan2022language,madaan2022language,wang2023codet5plus,hayati2018retrievalbased,beau2022impact,ahmed2024automatic}. In repository-level code completion, due to the massive amount of code in the repository and limited context of generator~\cite{ding2023cocomic}, it is impractical to use the entire repository as the context for generation. Therefore, most current methods employ the RAG method to retrieve suitable candidates from the repository for generation~\cite{ding2023cocomic,zhang2023repocoder,eghbali2024dehallucinator,zhang2024codeagent}.

\begin{figure}[ht]
\centering
\includegraphics[width=\linewidth]{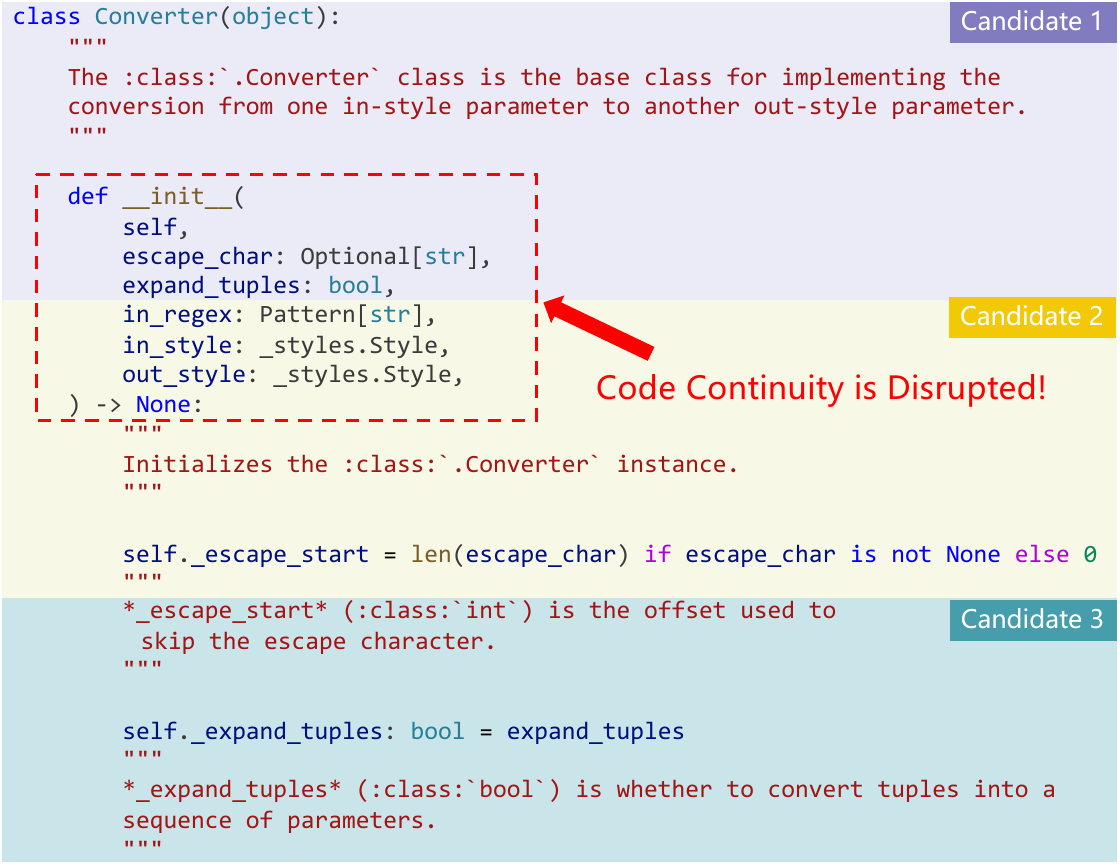}
\vspace{-5pt}
\caption{Using fixed window candidates may disrupt the continuity of code semantics, resulting in the definition of functions being split across different code snippets.}
\label{fixed_window_candidate}
\end{figure}

\begin{figure}[ht]
\centering
\includegraphics[width=\linewidth]{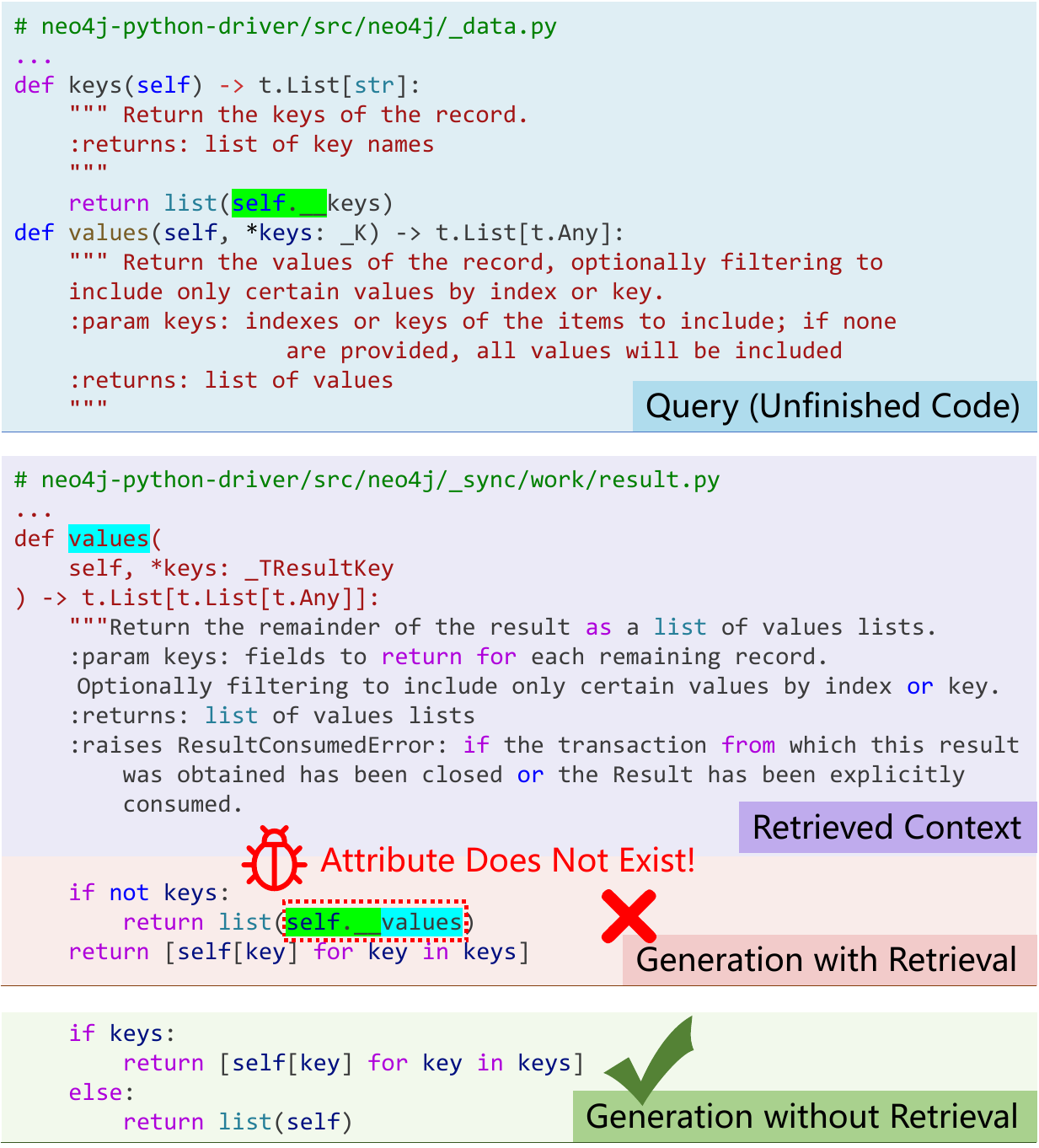}
\vspace{-5pt}
\caption{Due to the limitations of LLMs, inappropriate retrieval can mislead generation, resulting in attempts to call an non-existent attribute.}
\label{blind_retrieve}
\end{figure}

\subsection{Repository-Level Code Completion}
Traditional code completion~\cite{chen2021evaluating,austin2021program} usually focused on generating code with in-file context. With the development of LLMs~\cite{bubeck2023sparks,rozière2024code,li2023starcoder,guo2024deepseekcoder}, repository-level code completion is gradually gaining attention as it better reflects real-world scenarios~\cite{Yu2023CoderEvalAB,ding2023cocomic,zhang2023repocoder,ding2023crosscodeeval}. To formalize repository-level code completion, we conceptualize the process as selecting the most relevant snippets (candidates) from a code repository and generating code based on the query. This can be encapsulated in a formula as follows:
\[
\text{Code} = \text{Generate}(q, \text{Retrieve}(q, \text{codebase})) \tag{2}
\]
where \(q\) represents the query or the prompt for code completion. Codebase symbolizes code snippets from the code repository. \(\text{Retrieve}(q, \text{codebase})\) is the function that selects the most relevant code snippets (candidates) from the repository based on the query \(q\). \(\text{Generate}(q, \text{candidates})\) is the generation function that generates the target code based on the query \(q\) and the selected candidates.

Previous work highlights the importance of integrating both the in-file and cross-file context in repository-level code completion~\cite{ding2023cocomic}. This implies that the model needs to understand not only the local context but also third-party libraries and global modules~\cite{liao2024a3codgen}. Fusing analogy context and rationale context can greatly ensure the integrity of the retrieval codebase~\cite{liang2024repofuse}. Iterative retrieval and generation method~\cite{zhang2023repocoder,eghbali2024dehallucinator} involves concatenating the results generated from the previous iteration with the prior context to form a query. This query is then used for the subsequent round of retrieval and generation. Additionally, agents~\cite{zhang2024codeagent,hong2023metagpt,huang2024agentcoder,wang2024teaching} that assist in code completion through invoking tools or collaborating with each other is also a remarkable approach.

\noindent{\bf Limitations:} There are still some issues that need to be addressed for current repository-level code completion methods. \textit{First}, the lack of labeled data limits the generalizability of many learning-based approaches. For example, 
%the label (gold snippet)~\cite{liu2023repobench} for repository-level code completion tasks refers to the optimal context for prediction. Although there are some learning-based approaches for repository-level code completion tasks, the lack of labeled data limits the generalizability of these methods. 
CoCoMIC~\cite{ding2023cocomic} can only be used in trained repository and struggles to expand to other languages and repositories. RepoHyper~\cite{phan2024repohyper} uses a subset of the benchmark as training data and sets the gold candidate as label. 
\textit{Second}, previous works mostly adopted fixed window candidates\cite{zhang2023repocoder} or candidates based on dependency parsing~\cite{ding2023cocomic}. Methods based on dependency parsing only consider the nodes in the dependency parse graph, neglecting other code in the repository. This can lead to omitting many potentially useful code pieces during retrieval. Methods using fixed window candidates, as shown in Figure~\ref{fixed_window_candidate}, may split the signature of ``\texttt{\_\_init\_\_}'' function into two different candidates. This situation may lead to the retriever fetching the required code snippet without capturing the full parameter list. As a result, this partial information could confuse the generator leading to incorrect function calls. \textit{Third}, current work lacks an evaluation of the necessity for candidates. As illustrated in Figure~\ref{blind_retrieve}, the task can be correctly completed using only the in-file preceding context. However, if the context retrieved is blindly used, it may mislead the generation results. In this case, there is a function definition for ``\texttt{keys}'' in the unfinished code, which calls the ``\texttt{\_\_keys}'' attribute. The code snippet retrieved happens to have a function definition for ``\texttt{values}'', leading the model to mistakenly believe there is a corresponding ``\texttt{\_\_values}'' attribute defined, thus calling a non-existent ``\texttt{\_\_values}'' attribute during code generation.

\section{Methodology}
\subsection{Overview}

% \begin{figure*}[t]
% \centering
% \includegraphics[width=0.8\linewidth]{figure/overview.pdf}
% \caption{Overview of \model.}
% \label{Overview}
% \end{figure*}

\begin{figure}[t]
\centering
\includegraphics[width=\linewidth]{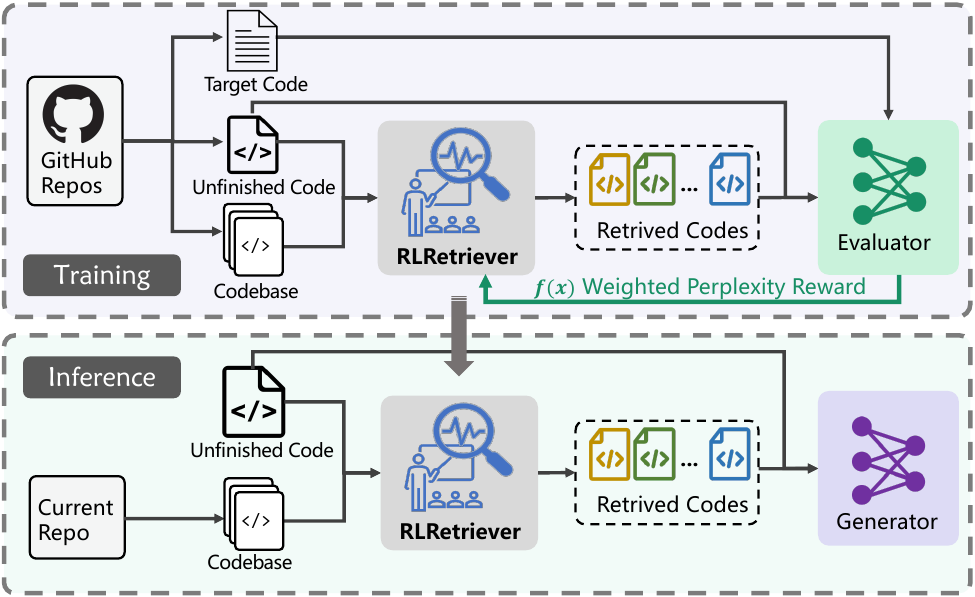}
\vspace{-15pt}
\caption{Overview of \model.}
\label{Overview}
\end{figure}

In this section, we introduce \model, a reinforcement learning framework for repository-level code completion. The overview of \model is shown in Figure~\ref{Overview}, comprising two stages: training and inference. 
In the training stage, the major objective is to train the retriever \rlretriever, the key component of our framework. First, to train \rlretriever, we construct data from repositories collected from GitHub and obtain unfinished code, target code, and candidate codebase. Then, \rlretriever will retrieve from the candidate codebase using unfinished code as the query. Finally, the retrieved code candidates will be evaluated by the evaluator and obtain the weight perplexity reward to update the parameters of \rlretriever. Through repeated iterations, \rlretriever enhances its retrieval capability via continuous feedback and learning. 
In the inference stage, given unfinished code and the current repository context, we first construct codebase from current repository. Then, we use the \rlretriever trained in the training stage to retrieve from the codebase using the unfinished code. Finally, we use the retrieved codes as context to concatenate with the unfinished code and feed them into the generator for target code generation.

\subsection{Data Construction}

\begin{figure}[t]
\centering
\includegraphics[width=0.45\textwidth]{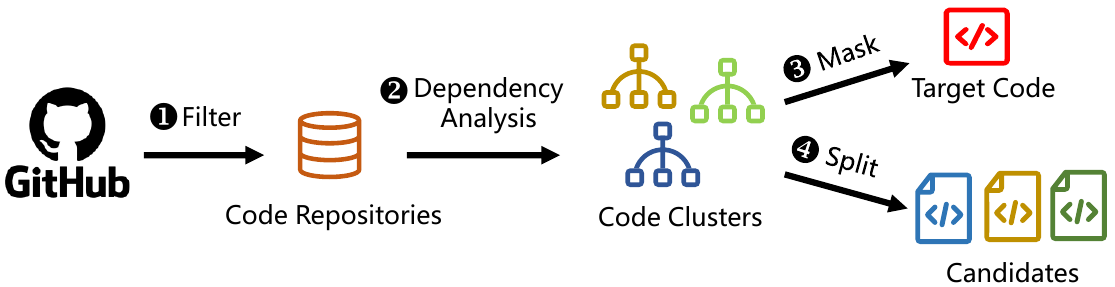}
\caption{Data construction pipeline.} %Training data construction pipeline.
\vspace{-5pt}
\label{Data_Construction}
\end{figure}

Repository-level code completion tasks typically refer to generating partial code within the given repository code context, generally including a line, an API, or a part of a function body~\cite{zhang2023repocoder}. To ensure the retriever we train meets the requirements of repository-level code retrieval, we need to simulate such scenarios. Figure~\ref{Data_Construction} shows the pipeline of our data construction process, which includes the following steps: repository filtering, dependency analysis, target code selection, and candidate construction. 

\subsubsection{Repository Selection} %Repository Filtering
% We randomly select 10,000 repositories from GitHub that meet the following requirements as our training dataset: (1) created before March 2023 (the release time of DeepSeekCoder~\cite{guo2024deepseekcoder}) so that xxx; (2) 
% several large-scale Python and Java projects from GitHub , explicitly ensuring these projects are created after the release of DeepSeekCoder~\cite{guo2024deepseekcoder} and not included in well-known benchmarks such as CrossCodeEval and RepoEval, which are used in our evaluation. This filtering process is aimed at preventing any data leakage, thereby ensuring the reliability of the evaluation. 

We randomly select 10,000 large-scale Python and Java repositories from GitHub that were created before March 2023 and meet the following requirements: (1) have cross-file dependencies for constructing our training dataset; and (2) not included in well-known benchmarks such as CrossCodeEval~\cite{ding2023crosscodeeval} and RepoEval~\cite{zhang2023repocoder}, which are used in our evaluation. This filtering process aims at preventing potential data leakage, ensuring the reliability of the evaluation. 

\subsubsection{Dependency Analysis}
To ensure our training data contains a substantial quantity of cross-file context dependencies, we implement a dependency analysis for each repository. The methodology is outlined in Algorithm~\ref{alg:Dependency_Analysis}. Specifically, to get the code files that are related to each other, we analyze the \texttt{import} statements within code files and construct a dependency graph that represents the relationships between these files. Based on whether dependencies exist between code files, we categorize them into clusters of interdependent code files. Through this process, we obtain 27,919 Python file clusters and 41,647 Java file clusters. We eliminate any cluster that contains only a single file. For clusters comprising multiple files, we employ a topological sorting based on the in-degree and out-degree of files. This means that, aside from the first file, each file in the cluster will contain code segments that depend on one or more of other files. 

\begin{algorithm}[t]
\caption{Dependency Analysis and Clustering.}
\label{alg:Dependency_Analysis}
\begin{algorithmic}
\REQUIRE Set of code files $F$
\ENSURE Clusters of interdependent code files $Clusters$

\STATE $G \leftarrow \text{ConstructDependencyGraph}(F)$
\STATE $Clusters \leftarrow \text{IdentifyClusters}(G)$
\FORALL{$cluster$ in $Clusters$}
    \IF{$\text{Size}(cluster) == 1$}
        \STATE $Clusters \leftarrow Clusters - \{cluster\}$
    \ELSE
        \STATE $SortedCluster \leftarrow \text{TopologicalSort}(cluster)$
        \STATE Update $cluster$ in $Clusters$ with $SortedCluster$
    \ENDIF
\ENDFOR
\RETURN $Clusters$
\end{algorithmic}
\end{algorithm}

\subsubsection{Target Code Selection}
Within the clusters of interdependent code files, we designate files other than the first file as the ones to be completed. We select a random position within these files, excluding the beginning and end to ensure ample context for the code to be completed. This position serves as the starting point for the target code segment that needs completion. To formalize this process, we define the target code segment to be masked and completed as \(C_{target}\), starting from position \(p_{start}\) with length \(l\), where \(p_{start}\) is chosen randomly within the constraints mentioned above. The selection of \(p_{start}\) can be expressed as:
\[
p_{start} = \text{Random}(p_{min}, p_{max}) \tag{3}
\]
where \(p_{min}\) and \(p_{max}\) define the permissible range within the file, excluding the very beginning and ending segments to ensure sufficient context. The length \(l\) of the target code \(C_{target}\) is also determined randomly, with the constraint that the entire segment \(C_{target}\) must lie within the boundary of the code file:
\[
C_{target} = C[p_{start} : p_{start} + l] \tag{4}
\]
After identifying \(C_{target}\), we \textit{mask} this segment within the file to simulate an unfinished code scenario that needs completion. Upon masking \(C_{target}\), the segment designated for completion, we intentionally exclude the file containing the masked code when assembling candidates. To formalize this concept, we define a binary selection function for candidate files as \(S(f_i)\), where \(f_i\) represents a candidate file:
\[
S(f_i) = 
\begin{cases} 
0 & \text{if } C_{target} \in f_i, \\
1 & \text{otherwise}
\end{cases} \tag{5}
\]

\subsubsection{Candidate Construction}
    
\begin{figure*}[t]
\centering
\includegraphics[width=\textwidth]{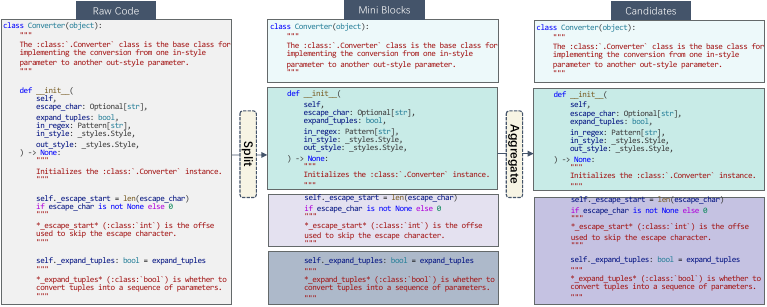}
\vspace{-5pt}
\caption{Candidate construction strategy.} %based on human programming habits
\label{candidate_construction}
\end{figure*}

Unlike previous works that utilized fixed window candidates~\cite{zhang2023repocoder,eghbali2024dehallucinator} or candidates parsed from dependencies~\cite{ding2023cocomic}, we propose a simple yet effective Split-Aggregate candidate construction strategy inspired by human programming habits. We term these candidates as natural candidates. Specifically, programmers often write code with continuous semantic information together, using blank lines as separators to facilitate readability. As shown in the left part of Figure~\ref{candidate_construction}, code and its corresponding comments are usually not separated by blank lines. In fact, blank lines are usually used to separate code snippets with different semantics and usage. This practice naturally forms continuous code segments. The Split-Aggregate strategy is outlined in Algorithm~\ref{alg:Split_Aggregate_Method}. Specifically, we divide the code in a file into several mini-blocks based on blank lines and then aggregate these mini-blocks into candidates by a certain length. During aggregation, the mini-blocks are concatenated to form candidates in such a way that the length of any candidate does not exceed a preset threshold value $T$. %We describe this process in detail in Algorithm~\ref{alg:Split_Aggregate_Method}.

\begin{algorithm}
\caption{Split-Aggregate Strategy}
\label{alg:Split_Aggregate_Method}
\begin{algorithmic}
\REQUIRE Code File $F$, Threshold $T$
\ENSURE Candidate set $C$
\STATE $Blocks \leftarrow \text{SplitIntoBlocks}(F)$
\STATE $C \leftarrow \emptyset$
\FORALL{$block$ in $Blocks$}
    \IF{$\text{LineCount}(block) < T$}
        \STATE $Aggregate \leftarrow block$
        \WHILE{$\text{LineCount}(Aggregate) < T$ \AND $block \neq \text{Last}(Blocks)$}
            \STATE $block \leftarrow \text{Next}(block)$
            \STATE $Aggregate \leftarrow Aggregate + block$
        \ENDWHILE
        \STATE $C \leftarrow C \cup \{\text{CreateCandidate}(Aggregate)\}$
    \ELSE
        \STATE $C \leftarrow C \cup \text{SplitBlock}(block, T)$
    \ENDIF
\ENDFOR
\RETURN $C$
\end{algorithmic}
\end{algorithm}
\vspace{-5pt}

\subsection{Reinforcement Learning-based \rlretriever Training}

\subsubsection{Design of Reward}

% \begin{figure}[t]
% \centering
% \includegraphics[width=0.8\linewidth]{figure/reward.pdf}
% \caption{Design of Reward.}
% \label{reward}
% \end{figure}

For reinforcement learning, reward is feedback from the external environment that assists a model in learning specific capabilities based on the feedback. In the scenario of repository-level code completion, the most intuitive indicator of reward is whether the generated code can be executed to obtain the expected results. However, obtaining feedback through actual execution is difficult. On the one hand, it's challenging to set up the execution environment for repository code. Even if the execution environment is established, execution can be time-consuming, and there may be a lack of corresponding test cases to evaluate the accuracy of the execution results.

In the context of repository-level code completion, our primary aim is to identify the optimal candidate \(c\) from a set of possibilities that maximizes the likelihood of accurately generating the target code sequence \(y\) given the contextual information \(x\). This objective can be formally articulated as:
\[
\max_{c} P(y|x, c) \tag{6}
\]
It is evident that this maximization is equivalent to minimizing the negative log-likelihood:
\[
\min_{c} -\log P(y|x, c) \tag{7}
\]
Perplexity (PPL), a standard measure for evaluating the predictive performance of probabilistic models, is defined as the exponential of the average negative log-likelihood (NLL) over a sequence. Minimizing NLL thereby directly corresponds to minimizing the perplexity of the target code sequence \(y\):
\[
\min_{c} PPL(y|x, c) = e^{-\frac{1}{N} \sum_{i=1}^{N} \log P(y_i|x, c, y_{<i})} \tag{8}
\]
where \( y_i \) represents the i-th token in the target code sequence \( y \).

In the domain of code completion, the first few tokens generated play a pivotal role in shaping the entire output. Considering this, we give more attention to the first few tokens. Besides, errors in repository-level code completion often occur due to hallucinations caused by a lack of understanding of the entire repository, such as generating incorrect or non-existent APIs. Therefore, we assign a higher focus on the identifier tokens. To further refine the model's focus, we introduce a weighted variant \(PPL_w\):
\[
PPL_w(y|x, c) = e^{-\frac{1}{\sum_{i=1}^{N} w_i} \sum_{i=1}^{N} w_i \cdot \log P(y_i|x, c, y_{<i})} \tag{9}
\]

\noindent The weight \(w_i\) for each token of the target code is determined by a function that considers the token's position in the sequence and whether it is an identifier, which can be represented as:
\[
w_i = \begin{cases} 
w_{first} & \text{if } i \leq k, \\
w_{api} & \text{if } y_i \in \text{APIs}, \\
1 & \text{otherwise}
\end{cases} \tag{10}
\]
where the first \(k\) tokens are assigned by a weight \(w_{first}\) to reflect their significant impact on the overall quality of the generated code. If the i-th token is part of an API or an identifier, it is assigned a weight \(w_{api}\) to acknowledge the importance of accurate and contextually
appropriate identifiers in code completion.

We define the reward for choosing a particular candidate \(c_i\), \(c_j\) from the set of all candidates \(C\) as follows:
\[
r(c_i) = 
\begin{cases} 
1 & \text{if } PPL_w(c_i) \leq PPL_w(c_j), \forall c_j \in C, \\
0 & \text{otherwise}
\end{cases} \tag{11}
\]
where \(PPL_w(c_i)\) denotes the weighted perplexity of the target code given candidate \(c_i\), serving as an abbreviation for \(PPL_w(y|x, c_i)\). The reward \(r(c_i)\), equivalently referred to as \(reward(c_i, x, C)\) in the formulations, is assigned a value of 1 if the candidate \(c_i\) exhibits a PPL that is equal to or lower than that of any other candidate in the set \(C\). Conversely, a reward of 0 is allocated to \(c_i\) if it fails to meet this criterion.

Building on the concept of this reward mechanism, we further define our objective function, \(\mathcal{L}\), as an aggregation of the logarithmic probabilities of choosing each candidate, weighted by the corresponding reward. This is mathematically represented as:
\[
\mathcal{L} = \sum_{i=1}^{n} \left( reward(c_i, x, C) \times \log p(c_i | x, C) \right) \tag{12}
\]
\noindent where \(n\) is the total number of candidates in set \(C\), and \(p(c_i | x, C)\) denotes the probability of selecting candidate \(c_i\) given the context \(x\) and the set of candidates \(C\).

\subsubsection{Stop Signal Mechanism for Candidates Selection}
\label{sec:Stop_Signal}

\begin{figure}[t]
\centering
\includegraphics[width=0.7\linewidth]{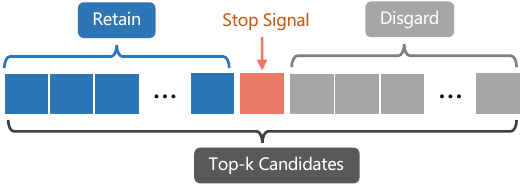}
\vspace{-2pt}
\caption{Illustration of the stop signal mechanism.} %Automatically Stop Retrieval
\label{stop}
\end{figure}

%Previous works mostly focus on when retrieval is needed~\cite{wang2024teaching}, overlooking which candidate to reserve. 
Previous works often overlook when to retrieve and which candidates to retain after retrieval. 
Specifically, after obtaining the top \(k\) candidates, traditional methods simply truncate this list to the top \(i\) candidates, determined by a predefined context length. However, this method overlooks the fact that not every retrieved candidate contributes positively to the generation process, and some may even have a negative impact. Therefore, discerning which candidates to retain is crucial for code completion performance. As shown in Figure~\ref{stop}, we design a stop signal mechanism. For each repository, this mechanism introduces an empty candidate serving as the stop signal into the candidate codebase. Within the candidate list retrieved by \rlretriever, we only retain those appearing before the stop signal. If the stop signal appears at the beginning of the list, it suggests that the generation task is likely to be a task that does not require cross-file context, as illustrated in Figure~\ref{blind_retrieve}.

\subsubsection{Learning from Reward}

As depicted in Figure~\ref{Overview}, \rlretriever is fine-tuned through a dynamic learning process where it receives rewards from the evaluator to update its parameters. This iterative learning process enables \rlretriever to progressively improve its retrieval results, leading to the selection of code candidates with progressively higher quality.

\subsection{Code Completion with \model}
%As shown in the lower half of Figure~\ref{Overview}, during the inference stage, we construct the candidate codebase from the current repository using the Split-Aggregate method. We retrieve code candidates using the unfinished code and the trained \rlretriever. Of course, we employ the stop strategy mentioned in Section~\ref{sec:Stop_Signal} to retain only the useful candidates. Subsequently, the unfinished code, together with these selected candidates, is provided as input to the generator for code completion.

As shown in the lower half of Figure~\ref{Overview}, during the inference stage, given unfinished code and the current repository context, we first construct the candidate codebase from the current repository using the Split-Aggregate method. Then, we retrieve code candidates using the unfinished code with the trained \rlretriever. Inherently, the stop signal strategy mentioned in Section~\ref{sec:Stop_Signal} is embedded in the retrieved results to retain only the useful candidates. Finally, the unfinished code, together with these selected candidates, is provided as input to the generator for code completion.

\section{Experimental Setup}
% In this section, we introduce the experimental setup, including baselines, benchmarks, evaluation metrics, and training details. The Research Questions are as follows:

% In this section, we present the metrics of code suggestions, the datasets we use in this paper, as well as the implementation details of our framework.

% We design experiments to verify the effectiveness of \model. Our evaluation focuses on the following research questions:

% \subsection{Experimental Settings}
\subsection{Baselines}~\label{sec:baselines}

%\noindent{\textbf{Evaluation for \model.}} 
\subsubsection{For \model}
To evaluate the effectiveness of \model, we compare it with the RawRAG method and RepoCoder framework. Besides, we use a popular dense retriever UniXcoder\cite{guo2022unixcoder} in these experiments.
\begin{itemize}
    \item \textbf{RawRAG} refers to the standard retrieval and generation approach in the repository-level code completion task. For the unfinished code to generate, RawRAG uses the left context of unfinished code as the query to find the relevant code in the repository to build prompts for generation.
    \item \textbf{RepoCoder}~\cite{zhang2023repocoder} is the state-of-the-art framework for repository-level code completion. It uses an iterative retrieval and generation approach to generate target code.
\end{itemize}

% \noindent{\textbf{Evaluation for \rlretriever.}} 
\subsubsection{For \rlretriever} 
To evaluate the effectiveness of \rlretriever, we compare it with the following commonly used retrieval methods in RAG:
\begin{itemize}
    \item \textbf{NoRetrieval} stands for direct generation with unfinished code, without retrieval. 
    \item \textbf{BM25}~\cite{robertson2009BM25} calculates scores for code candidates based on the frequency of query terms in each candidate. It adjusts for candidate length and the average candidate length across the entire database to prevent bias towards longer candidates. 
    \item \textbf{UniXcoder}~\cite{guo2022unixcoder} is a dense retriever that encodes both the query and the code snippets into dense vector spaces. This encoding facilitates the identification and retrieval of semantically relevant code snippets from a large corpus based on the similarity of vector representations.
    \item \textbf{UniXcoder-SFT} is a retriever that we trained using supervised fine-tuning of UniXcoder. Due to the lack of labeled data, we use the candidate with the lowest perplexity of target code as the label to fine-tune the retriever.
\end{itemize}

\subsection{Benchmarks}

\begin{table}[t]
    \centering
    \setlength{\tabcolsep}{2.7pt}
    % \begin{threeparttable}
    \caption{Benchmark statistics.}
    \vspace{-5pt}
    \label{tab:Benchmark}
    \begin{tabular}{ccccc}
    \toprule
    \textbf{Benchmark} & \textbf{Category} & \textbf{\#Samples} & \textbf{Avg. \#Lines} & \textbf{Avg. \#Tokens}\\ %\tnote{*}
    \midrule
    \multirow{2}{*}{CrossCodeEval} & Python & 2665 & 1.00 & 14.45 \\
    & Java & 2139 & 1.09 & 16.76 \\
    \cmidrule{1-5}
    \multirow{2}{*}{RepoEval} & Python (Line) & 1600 & 1.00 & 15.03 \\
    & Python (API) & 1600 & 2.48 & 34.91 \\
    % \cmidrule{1-5}
    % \multirow{2}{*}{GithubEval} & Python & 1000 & - & - \\
    % & Java & 1000 & - & - \\
    \bottomrule
    \end{tabular}
    % \begin{tablenotes}
    % \item[*] \textbf{Lines} and \textbf{Tokens} refer to the average number of lines and tokens of tasks within the benchmark, where the tokens are tokenized by the tokenizer of DeepSeekCoder-1B.
    % \end{tablenotes}
% \end{threeparttable}
\end{table}

We evaluate \model on widely used benchmarks for code completion: CrossCodeEval and RepoEval.
% Table~\ref{tab:Benchmark} provides detailed information on three benchmarks and an eval dataset named GithubEval, used to evaluate \model and \rlretriever. 
\begin{itemize}
\item \textbf{CrossCodeEval}~\cite{ding2023crosscodeeval} is a diverse and multilingual code completion benchmark, and we use the Python and Java parts of it.
\item \textbf{RepoEval}~\cite{zhang2023repocoder} is a benchmark proposed simultaneously with RepoCoder~\cite{zhang2023repocoder}. The benchmark consists of the latest repositories that cover the line-level, API-level, and function-level completion tasks. We use the line-level and API-level tasks among it for evaluation.
%\item \textbf{CoderEval}~\cite{Yu2023CoderEvalAB} is a function-level code completion benchmark. It provides test cases to verify the correctness of the generated code. We use the Python part of the benchmark to evaluate the ability to generate long code.
%\item \textbf{GitHubEval} is collected from GitHub and processed similarly to the training dataset. This dataset highlights code completion tasks at random positions within repositories.
\end{itemize}

Table~\ref{tab:Benchmark} shows the statistics of the benchmarks. \texttt{\#Samples} stands for the number of samples in a benchmark,  \texttt{Avg. \#Lines} and \texttt{Avg. \#Tokens} stands for the average numbers of lines and tokens of the target code snippets in the benchmark, respectively. Tokens are tokenized by the tokenizer of DeepSeekCoder-1B.

\subsection{Evaluation Metrics}

% For CrossCodeEval, RepoEval and GitHubEval, we use Exact Match and Edit Similarity as metrics. 
We measure the performance of our approach using the widely used metrics \textbf{\textit{Exact Match (EM)}} and \textbf{\textit{Edit Similarity (ES)}}~\cite{Levenshtein1965es}. These metrics are widely used in previous code completion studies~\cite{zhang2023repocoder, ding2023cocomic, ding2023crosscodeeval,eghbali2024dehallucinator}. 
EM assesses the precision of code completion by checking if the generated code matches the expected code exactly. It treats the entire code snippet as a single unit.
ES measures the similarity between the generated code and the expected code by calculating the edit distance. It reflects the number of edits needed to transform the generated code into the expected code.
    
% \begin{itemize}
%     \item \textbf{Exact Match (EM)} assesses the precision of code completion by checking if the generated code matches the expected code exactly. It treats the entire code snippet as a single unit.
%     \item \textbf{Edit Similarity (ES)}~\cite{Levenshtein1965es} measures the similarity between the generated code and the expected code by calculating the edit distance. It reflects the number of edits needed to transform the generated code into the expected code.
% \end{itemize}

\subsection{Experimental Details}
\label{sec:Experimental_Details}

% \begin{table}[!htbp]
%     \centering
%     \caption{Parameter Setting.}
%     \label{Parameter_Setting}
%     \begin{tabular}{ccc}
%     \hline
%     \textbf{Parameter Name} & \textbf{Setting} \\
%     \hline
%     epoch & 20 \\
%     batch size & 16 \\
%     sample number & 10 \\
%     data per epoch & 2000 \\
%     learning rate & 5e-5 \\
%     retriever query context length & 256 \\
%     retriever candidate context length & 256 \\
%     generator max crossfile length & 512 \\
%     generator max context length & 1024 \\
%     generator max generation length & 64 \\
%     \hline
%     \end{tabular}
% \end{table}

All experiments are conducted on a machine with two Tesla A100 GPUs, each with 80 GB memory. In the training stage, we use the parameters of UniXcoder~\cite{guo2022unixcoder} to initialize \rlretriever and use DeepSeekCoder-1B as the evaluator. The batch size is 16 and the learning rate is $5e^{-5}$. We train the model for 20 epochs with 2000 samples per epoch and perform early stopping. In the inference and evaluation stage, we use five different backbone models as the generators.
%In the training stage, we set the context length to 1024 (512 for in-file context and 512 for cross-file context). We use DeepSeekCoder-1B as the evaluator. In the inference stage, we expand the context length to 2048 (512 in-file + 1536 cross-file), and the generator to various backbone LLMs.

\section{Evaluation Results}
In this section, we report and analyze the experimental results to answer the following research questions (RQs):
\begin{itemize}
    \item \textbf{RQ1:} How effective is \model in repository-level code completion?
    
    \item \textbf{RQ2:} How effective is \rlretriever compared to other retrieval methods?
    % To assess the effectiveness of \rlretriever, we conduct a comparative analysis with several retrieval methods, including BM25~\cite{robertson2009BM25}, UniXcoder~\cite{guo2022unixcoder}, and supervised fine-tuned version of UniXcoder.
    
    \item \textbf{RQ3:} Does each component of \model contribute to its performance?
    % To evaluate the contribution of each component, We conduct an ablation study on \model. For the reward component, we use an unweighted perplexity of target code to train a retriever as a contrast. For NaturalCandidates, we use a fixed window candidate instead. Besides, we train a retriever disabling the stop signal to verify its contribution.

    \item \textbf{RQ4:} How is the generalizability of \model?
    % To assess the generalizability of \model, we conduct an evaluation with a setting different from the training stage. Additionally, we train a new retriever by fusing \model and RepoCoder. Then, we evaluate the performance of the fused model.
\end{itemize}

\subsection{RQ1: Effectiveness of \model}
% \subsection{RQ1: How effective is \model in repository-level code completion?}

\begin{table*}[t]
    \centering
    \setlength{\tabcolsep}{7pt}
    \caption{Performance of different models. The superscripts in percentage denote the improvement ratios of RLCoder over the corresponding best baseline. }
    %``Avg. Improvement'' denotes the mean of these improvement ratios under each metric.  } % on CrossCodeEval and RepoEval with 2048 context length, compared to RawRAG method and RepoCoder.
    \label{tab:effectiveness}
    \vspace{-5pt}
    % \scalebox{0.95}{
    \begin{tabular}{l ll ll ll ll}
    \toprule
    \multirow{2}{*}{\textbf{Model}} & \multicolumn{2}{c}{\textbf{CrossCodeEval (Python)}} & \multicolumn{2}{c}{\textbf{CrossCodeEval (Java)}} & \multicolumn{2}{c}{\textbf{RepoEval (Line)}} & \multicolumn{2}{c}{\textbf{RepoEval (API)}} \\
    \cmidrule(lr){2-3} \cmidrule(lr){4-5} \cmidrule(lr){6-7} \cmidrule(lr){8-9}
    & \textbf{EM} & \textbf{ES} & \textbf{EM} & \textbf{ES} & \textbf{EM} & \textbf{ES} & \textbf{EM} & \textbf{ES} \\
    \midrule
    \RawragCLSeven & 21.76 & 69.09 & 23.42 & 66.13 & 42.31 & 64.35 & 34.38 & 61.45 \\
    \RepocoderCLSeven & 23.34 & 70.84 & 24.17 & 66.56 & 43.94 & 65.81 & 37.00 & 63.51 \\
    \cellcolor{blue!5}\textbf{\RlcoderCLSeven} & 
    \cellcolor{blue!5}\textbf{26.60}\mytextsuperscript{ $ \uparrow$14.0\%}    & 
    \cellcolor{blue!5}\textbf{72.27}\mytextsuperscript{ $ \uparrow$2.0\%} & 
    \cellcolor{blue!5}\textbf{26.23}\mytextsuperscript{ $ \uparrow$8.5\%} & 
    \cellcolor{blue!5}\textbf{67.61}\mytextsuperscript{ $ \uparrow$1.6\%} & 
    \cellcolor{blue!5}\textbf{46.63}\mytextsuperscript{ $ \uparrow$6.1\%} & 
    \cellcolor{blue!5}\textbf{67.92}\mytextsuperscript{ $ \uparrow$3.2\%} & 
    \cellcolor{blue!5}\textbf{37.94}\mytextsuperscript{ $ \uparrow$2.5\%} & 
    \cellcolor{blue!5}\textbf{64.31}\mytextsuperscript{ $ \uparrow$1.3\%} \\
    \midrule

    \RawragSCSeven & 22.33 & 69.60 & 22.16 & 67.80 & 43.81 & 64.83 & 31.94 & 56.00 \\
    \RepocoderSCSeven & 23.15 & 70.71 & 22.53 & 68.22 & 45.69 & 66.90 & 33.44 & 57.81 \\
    \cellcolor{blue!5}\textbf{\RlcoderSCSeven} & 
    \cellcolor{blue!5}\textbf{25.82}\mytextsuperscript{ $ \uparrow$11.5\%} &
    \cellcolor{blue!5}\textbf{72.11}\mytextsuperscript{ $ \uparrow$2.0\%} & 
    \cellcolor{blue!5}\textbf{24.73}\mytextsuperscript{ $ \uparrow$9.8\%} & 
    \cellcolor{blue!5}\textbf{69.08}\mytextsuperscript{ $ \uparrow$1.3\%} & 
    \cellcolor{blue!5}\textbf{47.38}\mytextsuperscript{ $ \uparrow$3.7\%} & 
    \cellcolor{blue!5}\textbf{68.46}\mytextsuperscript{ $ \uparrow$2.3\%} & 
    \cellcolor{blue!5}\textbf{34.88}\mytextsuperscript{ $ \uparrow$4.3\%} & 
    \cellcolor{blue!5}\textbf{58.11}\mytextsuperscript{ $ \uparrow$0.5\%} \\
    \midrule

    \RawragSCTSeven & 22.89 & 70.66 & 23.42 & 69.13 & 44.44 & 65.95 & 34.50 & 58.78 \\
    \RepocoderSCTSeven & 24.35 & 71.71 & 23.75 & 69.59 & 45.81 & 67.37 & 36.44 & 59.92 \\
    \cellcolor{blue!5}\textbf{\RlcoderSCTSeven} & 
    \cellcolor{blue!5}\textbf{27.17}\mytextsuperscript{ $ \uparrow$11.6\%} & 
    \cellcolor{blue!5}\textbf{73.24}\mytextsuperscript{ $ \uparrow$2.1\%} & 
    \cellcolor{blue!5}\textbf{26.23}\mytextsuperscript{ $ \uparrow$10.4\%} & 
    \cellcolor{blue!5}\textbf{70.51}\mytextsuperscript{ $ \uparrow$1.3\%} & 
    \cellcolor{blue!5}\textbf{48.25}\mytextsuperscript{ $ \uparrow$5.3\%} & 
    \cellcolor{blue!5}\textbf{68.61}\mytextsuperscript{ $ \uparrow$1.8\%} & 
    \cellcolor{blue!5}\textbf{38.00}\mytextsuperscript{ $ \uparrow$4.3\%} & 
    \cellcolor{blue!5}\textbf{61.21}\mytextsuperscript{ $ \uparrow$2.2\%} \\
    \midrule

    \RawragDSCOne & 19.74 & 67.68 & 18.89 & 62.47 & 39.31 & 62.04 & 33.00 & 60.41 \\
    \RepocoderDSCOne & 20.23 & 68.78 & 19.59 & 62.35 & 40.88 & 63.56 & 35.13 & 61.92 \\
    \cellcolor{blue!5}\textbf{\RlcoderDSCOne} & 
    \cellcolor{blue!5}\textbf{23.98}\mytextsuperscript{ $ \uparrow$18.5\%} & 
    \cellcolor{blue!5}\textbf{70.44}\mytextsuperscript{ $ \uparrow$2.4\%} & 
    \cellcolor{blue!5}\textbf{20.80}\mytextsuperscript{ $ \uparrow$6.2\%} & 
    \cellcolor{blue!5}\textbf{63.39}\mytextsuperscript{ $ \uparrow$1.7\%} & 
    \cellcolor{blue!5}\textbf{44.19}\mytextsuperscript{ $ \uparrow$8.1\%} & 
    \cellcolor{blue!5}\textbf{66.48}\mytextsuperscript{ $ \uparrow$4.6\%} & 
    \cellcolor{blue!5}\textbf{36.06}\mytextsuperscript{ $ \uparrow$2.6\%} & 
    \cellcolor{blue!5}\textbf{62.72}\mytextsuperscript{ $ \uparrow$1.3\%} \\
    \midrule

    \RawragDSCSeven & 23.30 & 70.84 & 22.49 & 66.78 & 45.69 & 66.67 & 38.00 & 65.66 \\
    \RepocoderDSCSeven & 26.98 & 72.96 & 24.96 & 66.52 & 46.38 & 67.51 & 39.31 & \textbf{66.29} \\
    \cellcolor{blue!5}\textbf{\RlcoderDSCSeven} & 
    \cellcolor{blue!5}\textbf{30.28}\mytextsuperscript{ $ \uparrow$12.2\%} & 
    \cellcolor{blue!5}\textbf{74.42}\mytextsuperscript{ $ \uparrow$2.0\%} & 
    \cellcolor{blue!5}\textbf{26.09}\mytextsuperscript{ $ \uparrow$4.5\%} & 
    \cellcolor{blue!5}\textbf{67.31}\mytextsuperscript{ $ \uparrow$1.2\%} & 
    \cellcolor{blue!5}\textbf{48.75}\mytextsuperscript{ $ \uparrow$5.1\%} & 
    \cellcolor{blue!5}\textbf{69.43}\mytextsuperscript{ $ \uparrow$2.8\%} & 
    \cellcolor{blue!5}\textbf{39.88}\mytextsuperscript{ $ \uparrow$1.5\%} & 
    \cellcolor{blue!5}66.22\mytextsuperscript{ $ \uparrow$-0.1\%} \\
    % \midrule
    % \textbf{Avg. Improvement} & $\uparrow$13.56\% & $\uparrow$2.1\% & $\uparrow$7.88\% & $\uparrow$1.42\% & $\uparrow$5.66\% & $\uparrow$2.94\% & $\uparrow$3.04\% & $\uparrow$1.04\%\\
    \bottomrule
    \end{tabular}
    % }
\end{table*}

\begin{figure}[t]
\centering
\includegraphics[width=0.8\linewidth]{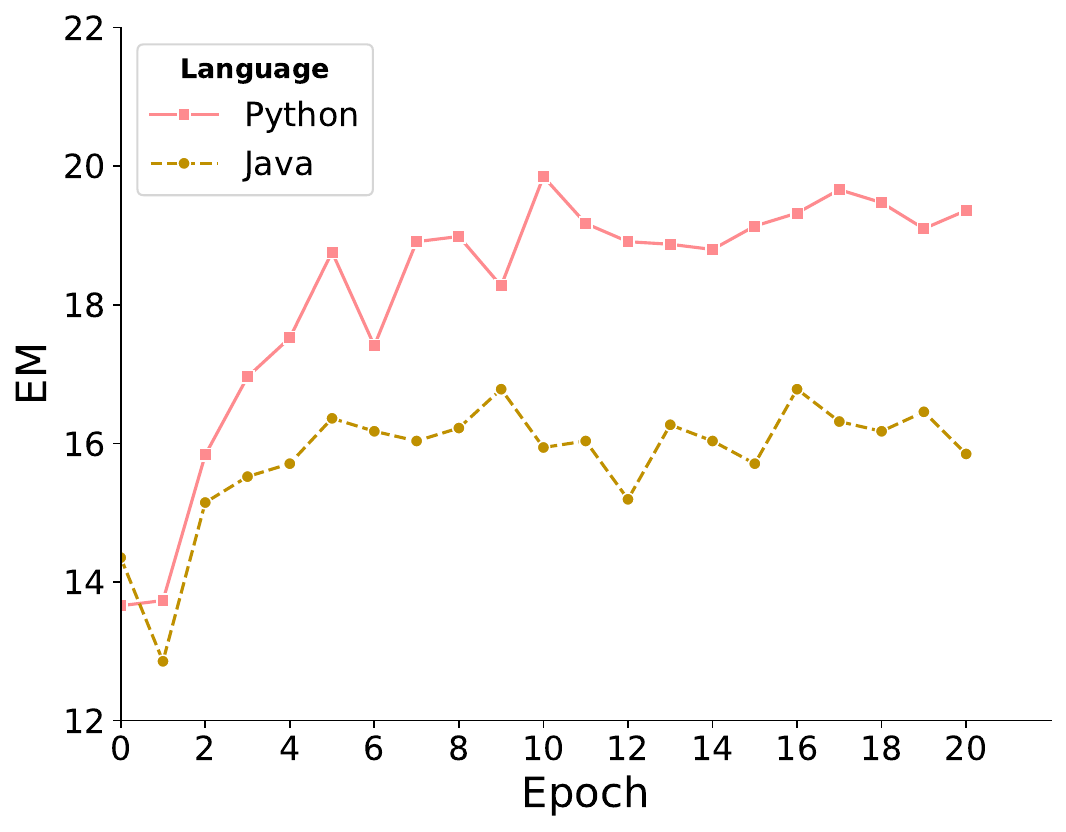}
\vspace{-5pt}
\caption{Performance trajectory curve during the training process.}
\label{effectiveness_epoch}
\end{figure}

To evaluate the effectiveness of \model, we compare it with the RawRAG framework~\cite{parvez2021rag} and RepoCoder~\cite{zhang2023repocoder} with five backbone LLMs, i.e., CodeLlama-7B~\cite{rozière2024code}, StartCoder-7B~\cite{li2023starcoder}, StarCoder2-7B, DeepSeekCoder-1B~\cite{guo2024deepseekcoder}, and DeepSeekCoder-7B. We evalaute the performance on CrossCodeEval~\cite{ding2023crosscodeeval} and RepoEval~\cite{zhang2023repocoder} benchmarks. 
%We evalaute the performance on four datasets, i.e., CrossCodeEval Python, CrossCodeEval Java, RepoEval Line Completion, and RepoEval API Completion. 
%with CrossCodeEval and RepoEval as our benchmarks.
%Table~\ref{tab:effectiveness} demonstrates the experimental results on the CrossCodeEval and RepoEval benchmarks, comparing \model's performance against RawRAG and RepoCoder. 

From the experimental results shown in Table~\ref{tab:effectiveness}, we can find that the proposed \model demonstrates effectiveness on all backbone language models across the four evaluated datasets, except for \RlcoderDSCSeven \  evaluated on RepoEval API, where its performance is on par with its corresponding best baseline \RepocoderDSCSeven. We can also observe that among all models, \RlcoderDSCSeven achieves the best performance with EM score of 30.28, improving its corresponding best baseline \RepocoderDSCSeven \  by 12.2\% on CrossCodeEval Python and 5.1\% on RepoEval Line.

% \begin{itemize}
%     \item Our proposed \model demonstrates effectiveness on all backbone language models across the four evaluated datasets. 
%     % outperforms the baseline methods RawRAG and RepoCoder with all backbone LLMs across various datasets. 
%     % For example, \RlcoderDSCOne improves its counterpart best baseline \RepocoderDSCOne by 18.5\%.   %when using DeepSeekCoder-1B, the EM of \model on the Python language of CrossCodeEval improves by approximately 21.48\% and 18.5\% compared to RawRAG and RepoCoder, respectively. 
%     Among all models, \RlcoderCLSeven achieves the best performance with EM score 30.28, improving its corresponding best baseline \RepocoderSCSeven by 12.2\%.
%     % \item We observed that \model shows slightly better performance in Python compared to Java. This phenomenon may be attributed to the fact that Java usually encapsulates less semantic information than Python with the same context length. 
%     % \item Furthermore, employing DeepSeekCoder-1B as the generation model yields a better result than baselines. This is likely due to the evaluator and generator having the same tokenizer. 
% \end{itemize}

Furthermore, to investigate the efficacy of our training process, we plot the performance trajectory curve across the training epochs on CrossCodeEval, as illustrated in Figure~\ref{effectiveness_epoch}. The result shows that the EM score gradually increases with each epoch until stabilizing, indicating the effectiveness of our training process.

\begin{center}
    % \begin{myboxb}[]{RQ1 Summary} %ab
    \begin{myboxc} \textbf{RQ1 Summary: } %cd
    Our approach significantly outperforms current state-of-the-art methods for all backbone LLMs, improving the CrossCodeEval benchmark by 12.2\% and RepoEval 5.1\%. The performance trajectory further demonstrates the efficacy of our training process.
    \end{myboxc} %cd
    % \end{myboxb} %ab
\end{center}

% From the experimental results shown in Table~\ref{tab:effectiveness}, we can find that the proposed \model demonstrates effectiveness on all backbone language models across the four evaluated datasets (except for \RlcoderDSCSeven evaluated on RepoEval API, where its performance is on par with its corresponding best baseline \RepocoderDSCSeven). We can also observe that among all models, \RlcoderDSCSeven achieves the best performance with EM score 30.28, improving its corresponding best baseline \RepocoderDSCSeven by 12.2\% on CrossCodeEval Python and 5.1\% on RepoEval Line.

\subsection{RQ2: Effectiveness of \rlretriever}
% \subsection{RQ2: How effective is \rlretriever compared to other retrieval methods?}

\begin{table*}[t]
    \centering
    \setlength{\tabcolsep}{7pt}
    \caption{Experimental results of \model equipped with different retrieval methods. The backbone LLM used is DeepSeekCoder-7B. The superscripts in percentage denote the improvement ratios of our retrieval model \rlretriever over the corresponding best baseline retriever.} %on CrossCodeEval and RepoEval compared to various retrieval methods
    \label{effectiveness_of_rlretriever}
    \begin{tabular}{lllllllll}
    \toprule
    \multirow{2}{*}{\textbf{Retrieval Method}} & \multicolumn{2}{c}{\textbf{CrossCodeEval (Python)}} & \multicolumn{2}{c}{\textbf{CrossCodeEval (Java)}} & \multicolumn{2}{c}{\textbf{RepoEval (Line)}} & \multicolumn{2}{c}{\textbf{RepoEval (API)}} \\
    \cmidrule(lr){2-3} \cmidrule(lr){4-5} \cmidrule(lr){6-7} \cmidrule(lr){8-9}
    & \textbf{EM} & \textbf{ES} & \textbf{EM} & \textbf{ES} & \textbf{EM} & \textbf{ES} & \textbf{EM} & \textbf{ES} \\
    \midrule
    \Baseline & 9.46 & 62.79 & 11.41 & 63.81 & 39.63 & 61.95 & 30.44 & 59.40 \\
    \BM & 18.31 & 68.38 & 17.48 & 65.23 & 45.94 & 66.67 & 38.25 & 65.04 \\
    \UniXcoder & 23.30 & 70.84 & 22.49 & 66.78 & 45.69 & 66.67 & 38.00 & 65.66 \\
    \UniXcoderSFT & 27.28 & 72.90 & 25.11 & 66.39 & 46.75 & 67.28 & 37.69 & 65.00 \\
    % \RLRetrieverSevenB & \textbf{30.28} & \textbf{74.42} & \textbf{26.09} & \textbf{67.31} & \textbf{48.75} & \textbf{69.43} & \textbf{39.88} & \textbf{66.22} \\
    \cellcolor{blue!5}\textbf{\RLRetriever} & \cellcolor{blue!5}\textbf{30.28}\mytextsuperscript{ $\uparrow$11.0\%} & \cellcolor{blue!5}\textbf{74.42}\mytextsuperscript{ $\uparrow$2.1\%} & \cellcolor{blue!5}\textbf{26.09}\mytextsuperscript{ $\uparrow$3.9\%} & \cellcolor{blue!5}\textbf{67.31}\mytextsuperscript{ $\uparrow$1.4\%} & \cellcolor{blue!5}\textbf{48.75}\mytextsuperscript{ $\uparrow$4.3\%} & \cellcolor{blue!5}\textbf{69.43}\mytextsuperscript{ $\uparrow$3.2\%} & \cellcolor{blue!5}\textbf{39.88}\mytextsuperscript{ $\uparrow$5.8\%} & \cellcolor{blue!5}\textbf{66.22}\mytextsuperscript{ $\uparrow$1.9\%} \\
    \bottomrule
    \end{tabular}
\end{table*}

To assess the effectiveness of \rlretriever, the key module of \model, we conduct a comparative study. We evaluate \model equipped with different retrieval methods described in Section~\ref{sec:baselines}, including NoRetrieval, BM25~\cite{robertson2009BM25}, UniXcoder~\cite{guo2022unixcoder}, and our enhanced model UniXcoder-SFT. Table~\ref{effectiveness_of_rlretriever} shows the experimental results on the CrossCodeEval and RepoEval benchmarks. The results yield the following findings:
\begin{itemize}
    \item Our proposed \rlretriever consistently outperforms comparative baseline methods under all metrics in both benchmarks, underscoring its superior performance. 
    \item All retrieval-based methods (i.e., BM25, UniXCoder, UniXcoder-SFT, and \rlretriever) perform better than NoRetrieval, showing the inherent value of the retrieval process itself.
    \item Both UniXcoder-SFT and \rlretriever show better performance than \UniXcoder, indicating that retrieval training can enhance the performance. Notably, our reinforcement learning-based training method exhibits better performance over supervised fine-tuning. % under the labelled data lacking constraint.
\end{itemize}

%We find that \rlretriever consistently outperforms other methods on all metrics of both benchmarks. For example, in the setting with a context length of 1024 and a generator of DeepSeekCoder-1B, \rlretriever's EM score is improved by 234\% compared to the baseline, by 47.16\% compared to UniXcoder, and by 26.84\% compared to supervised fine-tuned UniXcoder. We observe that BM25 underperforms, likely because it primarily assesses the similarity based on the frequency of tokens between the query and the candidates, without considering the deeper semantic information. UniXcoder and UniXcoder-SFT focus on the similarity between the query and candidates, overlooking the potentially truly desired candidate. In contrast, \rlretriever, which is trained to meet the generator's needs, can identify candidates that are beneficial for generation and successfully retrieve them.

\begin{center}
    % \begin{myboxb}[]{RQ2 Summary} %ab
    \begin{myboxc} \textbf{RQ2 Summary: } %cd
    % \begin{myboxd} \textbf{RQ2 Summary: } %cd
    \rlretriever consistently outperforms other retrieval methods. Furthermore, the results affirms the significance of the retrieval and training processes, particularly highlighting the advantages of our reinforcement learning-based training approach. 
    \end{myboxc} %cd
    % \end{myboxd} %cd
    % \end{myboxb} %ab
\end{center}

\subsection{RQ3: Contributions of Each Component}

\begin{table*}[t]
    \centering
    \setlength{\tabcolsep}{7pt}
    \begin{threeparttable}
    % \caption{Ablation Study on CrossCodeEval and RepoEval.}
    \caption{Ablation study results on CrossCodeEval and RepoEval.}
    \label{Ablation_Study}
    \begin{tabular}{lllllllll}
    \toprule
    \multirow{2}{*}{\textbf{Model}} & \multicolumn{2}{c}{\textbf{CrossCodeEval (Python)}} & \multicolumn{2}{c}{\textbf{CrossCodeEval (Java)}} & \multicolumn{2}{c}{\textbf{RepoEval (Line)}} & \multicolumn{2}{c}{\textbf{RepoEval (API)}} \\
    \cmidrule(r){2-3} \cmidrule(r){4-5} \cmidrule(r){6-7} \cmidrule(r){8-9}
    & \textbf{EM} & \textbf{ES} & \textbf{EM} & \textbf{ES} & \textbf{EM} & \textbf{ES} & \textbf{EM} & \textbf{ES} \\
    \midrule
    RLCoder & 30.28 & 74.42 & 26.09 & 67.31 & 48.75 & 69.43 & 39.88 & 66.22 \\
    \ \ w/o RL & 23.30\mytextsuperscript{ $\downarrow$23.1\%} & 70.84\mytextsuperscript{ $\downarrow$4.8\%} & 22.49\mytextsuperscript{ $\downarrow$13.8\%} & 66.78\mytextsuperscript{ $\downarrow$0.8\%} & 45.69\mytextsuperscript{ $\downarrow$6.3\%} & 66.67\mytextsuperscript{ $\downarrow$4.0\%} & 38.00\mytextsuperscript{ $\downarrow$4.7\%} & 65.66\mytextsuperscript{ $\downarrow$0.8\%} \\
    \ \ w/o WP & 27.35\mytextsuperscript{ $\downarrow$9.7\%} & 72.82\mytextsuperscript{ $\downarrow$2.1\%} & 25.67\mytextsuperscript{ $\downarrow$1.6\%} & 67.43\mytextsuperscript{ $\uparrow$0.2\%} & 47.44\mytextsuperscript{ $\downarrow$2.7\%} & 67.83\mytextsuperscript{ $\downarrow$2.3\%} & 38.81\mytextsuperscript{ $\downarrow$2.7\%} & 65.25\mytextsuperscript{ $\downarrow$1.5\%} \\
    \ \ w/o NC & 29.31\mytextsuperscript{ $\downarrow$3.2\%} & 73.91\mytextsuperscript{ $\downarrow$0.7\%} & 24.03\mytextsuperscript{ $\downarrow$7.9\%} & 66.49\mytextsuperscript{ $\downarrow$1.2\%} & 47.13\mytextsuperscript{ $\downarrow$3.3\%} & 68.11\mytextsuperscript{ $\downarrow$1.9\%} & 38.63\mytextsuperscript{ $\downarrow$3.1\%} & 65.56\mytextsuperscript{ $\downarrow$1.0\%} \\
    \ \ w/o SS & 29.57\mytextsuperscript{ $\downarrow$2.34\%} & 74.49\mytextsuperscript{ $\uparrow$0.09\%} & 25.57\mytextsuperscript{ $\downarrow$1.99\%} & 67.42\mytextsuperscript{ $\uparrow$0.16\%} & 47.31\mytextsuperscript{ $\downarrow$2.95\%} & 68.23\mytextsuperscript{ $\downarrow$1.73\%} & 39.63\mytextsuperscript{ $\downarrow$0.63\%} & 65.87\mytextsuperscript{ $\downarrow$0.53\%} \\
    \bottomrule
    \end{tabular}
    % \begin{tablenotes}
    % \item[*] \textit{w/o RL} (Reinforcement Learning) means using retriever without reinforcement learning. \textit{w/o WP} (Weighted Perplexity) means utilizing unweighted perplexity of the target code as the reward. \textit{w/o NC} (NaturalCandidates) signifies using fixed window candidates instead of NaturalCandidates.
    % \end{tablenotes}
    \end{threeparttable}
\end{table*}

To understand the contributions of each component to \model, we conduct an ablation study on \model. 
Specifically, we remove each component of \model each time and study the performance of the ablated model. The experimental results are shown in Table~\ref{Ablation_Study}. ``w/o RL'' means using retriever without reinforcement learning. ``{w/o WP'' means utilizing unweighted perplexity of the target code as the reward. ``w/o NC'' means using fixed window candidates instead of our natural candidates, ``w/o SS'' means using retriever without the stop signal mechanism. From Table~\ref{Ablation_Study}, we can see that the performance of the model drops after removing any one component, indicating that each component contributes to the effectiveness of \model. %an important role in code completion.  %We observe that the average score of EM decreases by approximately 3.5\% when we use the unweighted perplexity of the target code. Without NaturalCandidates, the average score of EM drops 10\%, which indicates that continuous code blocks are beneficial for generation. 
Especially, the performance drops the most significantly for ``RLCoder w/o RL'', indicating that the reinforcement learning mechanism is the most important component in \model. We observe that unweighted perplexity and no stop mechanism can be beneficial to ES performance in some cases, but still harm EM performance. In fact, ES mainly considers the similarity between two pieces of code. The introduction of the stop signal and weighted perplexity can both affect code similarity. The stop signal reduces useless but similar candidates, while weighted perplexity emphasizes API tokens more rather than all tokens. This can potentially reduce the similarity between generated and target code. Although these strategies weaken similarity, they improve code correctness. So removing the stop signal and weighted perplexity decreases in EM across all benchmarks.

% \begin{figure}[t]
% \centering
% \includegraphics[width=0.65\linewidth]{figure/Ablation_Stop.pdf}
% \caption{Ablation study of the stop signal mechanism.} % on GitHubEval.
% \label{Ablation_Stop}
% \end{figure}

\begin{table}[t]
    \centering
    \setlength{\tabcolsep}{7pt}
    \caption{Additional ablation study for the stop signal mechanism. Note that, contrary to EM and ES, lower PPL scores correspond to better performance.} % We set a limited context length, where in-file context length is 256 and cross-file context length is 768.
    %``Avg. Improvement'' denotes the mean of these improvement ratios under each metric.  } % on CrossCodeEval and RepoEval with 2048 context length, compared to RawRAG method and RepoCoder.
    \label{tab:Supplement_Ablation_Study}
    \vspace{-5pt}
    % \scalebox{0.95}{
    \begin{tabular}{l lll}
    \toprule
    \textbf{Model} & \textbf{EM} & \textbf{ES} & \textbf{PPL} \\
    \midrule
    \RawragCLSeven & 10.3 & 65.2 & 2.1389 \\
    \RlcoderCLSeven & 11.1 & 66.4 & 2.0835 \\
    \ \ w/o Stop Signal & 9.9\mytextsuperscript{ $\downarrow$10.81\%} & 66.1\mytextsuperscript{ $\downarrow$0.45\%} & 2.1152\mytextsuperscript{ $\uparrow$1.52\%} \\
    \midrule

    \RawragSCSeven & 8.9 & 59.2 & 2.7400 \\
    \RlcoderSCSeven & 10.1 & 61.3 & 2.6695 \\
    \ \ w/o Stop Signal & 9.1\mytextsuperscript{ $\downarrow$9.90\%} & 60.1\mytextsuperscript{ $\downarrow$1.96\%} & 2.7100\mytextsuperscript{ $\uparrow$1.52\%} \\
    \midrule

    \RawragSCTSeven & 9.1 & 60.5 & 2.6422 \\
    \RlcoderSCTSeven & 10.3 & 60.2 & 2.5716 \\
    \ \ w/o Stop Signal & 9.3\mytextsuperscript{ $\downarrow$9.71\%} & 60.5\mytextsuperscript{ $\uparrow$0.50\%} & 2.6115\mytextsuperscript{ $\uparrow$1.55\%} \\
    \midrule

    \RawragDSCOne & 9.6 & 64.9 & 2.4370 \\
    \RlcoderDSCOne & 10.5 & 66.7 & 2.3764 \\
    \ \ w/o Stop Signal & 10.0\mytextsuperscript{ $\downarrow$4.76\%} & 65.9\mytextsuperscript{ $\downarrow$1.20\%} & 2.4138\mytextsuperscript{ $\uparrow$1.57\%} \\
    \midrule

    \RawragDSCSeven & 11.5 & 66.6 & 2.3334 \\
    \RlcoderDSCSeven & 12.2 & 68.6 & 2.2824 \\
    \ \ w/o Stop Signal & 11.8\mytextsuperscript{ $\downarrow$3.28\%} & 67.9\mytextsuperscript{ $\downarrow$1.02\%} & 2.3157\mytextsuperscript{ $\uparrow$1.46\%} \\
    \bottomrule
    \end{tabular}
    % }
\end{table}

% To evaluate the effectiveness of the stop signal, we conduct a series of experiments on the GitHubEval. The results are shown in Figure~\ref{Ablation_Stop}. Under the setting of a fixed in-file context length of 256, we extend the cross-file context length from 256 to 2048 and calculate the corresponding EM and PPL. The results indicate that under limited context length, the stop signal can balance discarding less useful cross-file context to gain more in-file context to help generation. It is worth noting that when the context length is 2048, there is an intersection of curves between using and not using the stop signal. This occurs because the in-file context has been fully utilized at this point. Methods that do not use the stop signal actually have a longer context length at this point. In other words, methods using the stop signal achieve the same effect with less context.

Since CrossCodeEval and RepoEval are specifically curated to evaluate code completion ability in scenarios requiring cross-file context, the improvement brought by the stop signal is expected to be minor. To further evaluate the practical effectiveness of the stop signal mechanism, we construct a new dataset GitHubEval that construct code completion targets at random positions within repositories, thus incorporating instances that may not necessitate cross-file context. Following the same construction procedure as the training dataset, we obtain 1000 samples for evaluation. The results shown in Table~\ref{tab:Supplement_Ablation_Study} indicate that the stop signal is an important component in \model, without which, the EM scores drop significantly by an average of 7.69\%.

% To evaluate the effectiveness of the stop signal, we conduct a series of experiments on the GitHubEval. The results are shown in Figure~\ref{Ablation_Stop}. Under the setting of a fixed in-file context length of 256, we extend the cross-file context length from 256 to 2048 and calculate the corresponding EM and PPL. The results indicate that under limited context length, the stop signal can balance discarding less useful cross-file context to gain more in-file context to help generation. It is worth noting that when the context length is 2048, there is an intersection of curves between using and not using the stop signal. This occurs because the in-file context has been fully utilized at this point. Methods that do not use the stop signal actually have a longer context length at this point. In other words, methods using the stop signal achieve the same effect with less context.

\begin{center}
    % \begin{myboxb}[]{RQ3 Summary} %ab
    \begin{myboxc} \textbf{RQ3 Summary: } %cd
    % \begin{myboxd} \textbf{RQ4 Summary: } %cd
    Reinforcement learning mechanism is the most important component in \model. Other components of \model also contributes to its superior performance, with the stop signal mechanism showing further enhancements in scenarios involving target code completion that both require and do not require cross-file context.
    %Using weighted perplexity of the target code is important for \model, and natural candidates are a better choice compared to fixed window candidates. The stop signal can discard useless cross-file context to retain more in-file context when the in-file context is not sufficient.
    \end{myboxc} %cd
    % \end{myboxd} %cd
    % \end{myboxb} %ab
\end{center}

% \vspace{-1pt}
\subsection{RQ4: Generalizability of \model}
\vspace{5pt}

\begin{table*}[t]
    \centering
    \setlength{\tabcolsep}{8pt}
    \caption{Experimental results of RepoCoder integrated with \model.} %Experimental results with the setting of 2048 context length and use DeepSeekCoder-7B as the generator to evaluate the fusion of RepoCoder and \model.
    \label{Orthogonality}
    \vspace{-5pt}
    \begin{tabular}{lllllllll}
    \toprule
    \multirow{2}{*}{\textbf{Method}} & \multicolumn{2}{c}{\textbf{CrossCodeEval (Python)}} & \multicolumn{2}{c}{\textbf{CrossCodeEval (Java)}} & \multicolumn{2}{c}{\textbf{RepoEval (Line)}} & \multicolumn{2}{c}{\textbf{RepoEval (API)}} \\
    \cmidrule(r){2-3} \cmidrule(r){4-5} \cmidrule(r){6-7} \cmidrule(r){8-9}
    & \textbf{EM} & \textbf{ES} & \textbf{EM} & \textbf{ES} & \textbf{EM} & \textbf{ES} & \textbf{EM} & \textbf{ES} \\
    \midrule
    RawRAG & 23.30 & 70.84 & 22.49 & 66.78 & 45.69 & 66.67 & 38.00 & 65.66 \\
    RLCoder & 30.28 & 74.42 & 26.09 & 67.31 & 48.75 & 69.43 & 39.88 & 66.22 \\ \hdashline
    RepoCoder & 26.98 & 72.96 & 24.96 & 66.52 & 46.38 & 67.51 & 39.31 & 66.29 \\
    \ \ \ \cellcolor{blue!5}w/ \model & \cellcolor{blue!5}\textbf{30.32}\mytextsuperscript{ $\uparrow$12.4\%} & \cellcolor{blue!5}\textbf{74.79}\mytextsuperscript{ $\uparrow$2.5\%} & \cellcolor{blue!5}\textbf{26.98}\mytextsuperscript{ $\uparrow$8.1\%} & \cellcolor{blue!5}\textbf{67.81}\mytextsuperscript{ $\uparrow$1.9\%} &
    \cellcolor{blue!5}\textbf{49.44}\mytextsuperscript{ $\uparrow$6.6\%} & \cellcolor{blue!5}\textbf{69.76}\mytextsuperscript{ $\uparrow$3.3\%} & \cellcolor{blue!5}\textbf{41.25}\mytextsuperscript{ $\uparrow$4.9\%} & \cellcolor{blue!5}\textbf{67.08}\mytextsuperscript{ $\uparrow$1.2\%} \\
    \bottomrule
    \end{tabular}
\end{table*}

\begin{figure*}[t]
\centering
\includegraphics[width=\linewidth]{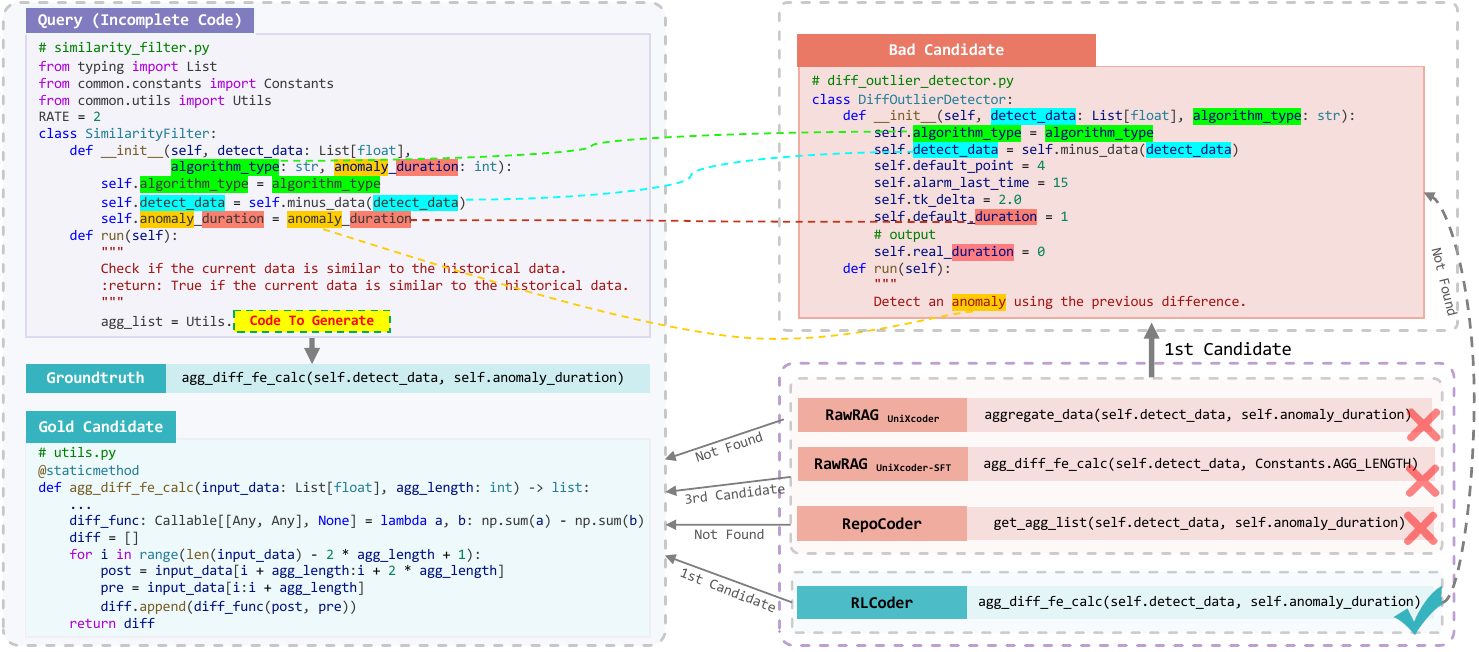}
\vspace{-20pt}
\caption{Case study. An example sources from CrossCodeEval with the \textit{task\_id} being \textit{project\_cc\_python/210}. The highlight token represents the identical tokens exists in both the query and the bad candidate.} %We use a context length of 2048 and DeepSeekCoder-7B as the generation model for this case. The highlight token represents the same token between the query and the bad candidate.
\label{Case_Study}
\end{figure*}
\vspace{-10pt}

%As mentioned in Section~\ref{sec:Experimental_Details}, we train the \rlretriever exclusively in a setting with a context length of 1024 and use DeepSeekCoder-1B as the evaluator. In the inference phase, we extend the context length to 2048 and adapt the generator to other backbones. From Table~\ref{effectiveness_of_rlcoder} and Table~\ref{effectiveness_of_rlretriever}, it is evident that our \rlretriever can adapt to other context lengths and backbone generation models and achieves remarkable performance. 

To explore the generalizability of \model, we conduct an evaluation with a setting different from the training stage. Specifically, we train a new retriever by fusing \model and RepoCoder. Then, we evaluate the performance of the fused model. 
% We apply the training framework of \model to RepoCoder. 
As shown in Table~\ref{Orthogonality}, we find that RepoCoder trained using the framework of \model significantly outperforms the original RepoCoder method. Specifically, the improvement rates in EM for Python and Java are 12.4\% and 8.1\% on CrossCodeEval, respectively. This result indicates that our training framework can be integrated into other models to further improve their performance. 
Note that when comparing ``RepoCoder w/ RLCoder'' to \model, they have comparable performance (with ``RepoCoder w/ RLCoder'' slightly better). However, considering that RepoCoder requires multiple iterations of retrieval and generation, we opt for RLCoder, which accomplishes code completion in a single round, as the default setting in this work.

% Furthermore, in terms of candidate construction, it is conceivable that our Split-Aggregate method might be replaceable by other candidate construction methods, such as those based on dependency analysis~\cite{ding2023cocomic,liang2024repofuse,phan2024repohyper}. Despite this, we believe our reinforcement learning framework could still remain effective, although this is yet to be empirically validated.

\begin{center}
    % \begin{myboxb}[]{RQ4 Summary} %ab
    \begin{myboxc} \textbf{RQ4 Summary: } %cd
    % \begin{myboxd} \textbf{RQ4 Summary: } %cd
    The training pipeline of \model shows generalizability on all datasets in applying to other frameworks.
    %\model shows generalizability in applying to other models. A trained \rlretriever can adapt to other context lengths and backbone generation models. Besides, the training framework of \model can significantly enhance RepoCoder.
    \end{myboxc} %cd
    % \end{myboxd} %cd
    % \end{myboxb} %ab
\end{center}

\vspace{-3pt}
\subsection{Case Study}
\vspace{-3pt}

% We demonstrate the effectiveness of \model through a case study. When evaluating on a task of CrossCodeEval, we find that other methods incorrectly rank a bad candidate as the first candidate, ultimately leading to generating code with the wrong call of API, as shown in Figure~\ref{Case_Study}. However, \model ranks the gold candidate as the first candidate and generates the target code correctly. A possible reason other methods fail to retrieve the correct code is that these methods focus too much on the similarity between the query and the candidate. However, the bad candidate just right has many same tokens as in the query, as highlighted in the query block and bad candidate block. Although RepoCoder can use the generated code to retrieve and generate iteratively, it still cannot avoid relying on the similarity between candidate and query for retrieval. However, \model considers the perplexity of the target code to generate given a certain candidate, thus allowing \rlretriever to avoid \textit{seemingly useful but actually useless} candidates, and prioritize candidates that may be helpful for generating. That is why \model successfully retrieves the gold candidate and correctly generates the right target code.

We illustrate the effectiveness of \model through a case study presented in Figure~\ref{Case_Study}. The left-hand side shows the incomplete code, groundtruth code, and the gold candidate we labelled for this case. 
We can see that \model ranks the gold candidate as the first candidate and generates the correct code. A possible reason that other methods fail to retrieve the correct code is that these methods rely on the surface-level similarity between the query and the candidate. The bad candidate, despite sharing several tokens with the query (as highlighted in the figure), does not contribute meaningfully to the correct code completion. 
% Although RepoCoder can iteratively use the generated code to retrieve and generate, it still cannot avoid relying on the similarity between candidate and query for retrieval. In contrast, \model leverages the perplexity of the target code with a given candidate, thus enabling \rlretriever to bypass candidates that are \textit{seemingly useful but actually useless} and prioritize those more likely to aid in accurate code completion. 
In contrast to RepoCoder, which iteratively uses generated code for retrieval and generation but still depends heavily on query-candidate similarity, \model leverages the perplexity of generating target code from a given candidate. This approach enables \rlretriever to bypass candidates that are \textit{seemingly useful but actually useless}, focusing instead on those more likely to aid in accurate code generation. This strategic prioritization explains \model's success in both retrieving the gold candidate and generating the correct target code.

% \section{Discussion}

% \section{Threats to Validity}

% \noindent \textit{\textbf{Internal Validity.}}
% Due to the limited computational resources, we train \rlretriever under a single setting (xxx). The optimal training settings remain unknown, but we have demonstrated that our trained models are effective across various context lengths and backbone LLMs. Another potential threat to internal validity is that due to the complexity of dependency parsing and the lack of relevant open-source code, we have only compared our natural candidates with fixed window candidates. In our experiments, we ignore the right context of the target code to do code completion. The effectiveness of \model in code-filling tasks is worth exploring.

% \noindent \textit{\textbf{External Validity}.}
% A potential threat to external validity is that we only evaluate \model on Python and Java datasets. Since Python and Java are two of the most popular languages, the evaluation results can still demonstrate the effectiveness of our method. Besides, performance may vary on different GPU architectures, slightly affecting \model's experimental results. In this paper, we choose the widely used architectures to reduce the basis.  

\vspace{-1pt}
\section{Related Work}
\vspace{-1pt}
\subsection{Code Completion}
\vspace{-1pt}

Code completion as one of the most important tasks in modern IDEs, has attracted the attention of many researchers~\cite{liu2021opportunities,chen2024identifying,wang2024beyond,wang2021code,tao2024kadel}. 
Traditional studies~\cite{bruch2009learning,hou2010towards,robbes2008program} use rule-based methods or code examples for code completion.
In recent years, deep learning-based methods~\cite{chen2023api,wang2019learning,liu2019generating,wang2022compilable,izadi2022codefill,zhou2022improving,wang2021code,aye2020sequence,hellendoorn2017deep,karampatsis2020big,kim2021code,li2017code,nguyen2019combining,svyatkovskiy2020intellicode,wen2021siri,yang2017language} have been explored to improve the performance of code completion. 
Recent studies found that code search can enhance code completion performance~\cite{chen2024code}. With the development of large language models~\cite{lu2024yoda,hu2022split,liu2024empirical,wang2024sparsecoder,zhou2023adaptive,shi2023sotana,zheng2023survey,zheng2023towards,chen2023chatgpt,zhong2023agieval}, many researchers have introduced LLMs into code completion~\cite{liu2023recommending,jiang2024seed,yang2024improving,li2024ircoco,li2023large,zhu2023improving,guo2022codefast}. Equipped with LLMs, many studies have employed RAG for code completion/generation~\cite{parvez2021rag,zan2022language,zhou2023docpromptingrag,li2023acecoder,hayati2018retrievalbased,zhang2023syntax}. For example, RedCoder~\cite{parvez2021rag} enhances code generation and summarization by integrating relevant past work using dense retrieval techniques. To enhance private library code generation, APICoder~\cite{zan2022language} was proposed to employ API documentation to train models to better generate these libraries. DocPrompting~\cite{zhou2023docpromptingrag} introduces a method to enhance code generation by using code documentation to address the challenge of generating code for unseen functions and libraries. AceCoder~\cite{li2023acecoder} improves code generation by integrating example retrieval and guided generation. ReCode~\cite{hayati2018retrievalbased} improves neural code generation by incorporating subtree retrieval from existing code examples. kNN-TRANX~\cite{zhang2023syntax} improves code generation from natural language by using syntax-aware retrieval, reducing noise and computational time.

% \textbf{\textit{Repository-Level Code Completion.}}    

\vspace{-2pt}
\subsection{Repository-Level Code Completion}
\vspace{-2pt}
 %wu2024repoformer
Repository-level code completion, which leverages the broader context of an entire code repository, has become a focal point for research in the field of code completion and many studies have attempted to improve repository-level code completion performance~\cite{ding2023cocomic,phan2024repohyper,bairi2023codeplan,liang2024repofuse,liao2024a3codgen,zhang2023repocoder,eghbali2024dehallucinator,zhang2024codeagent,wang2024teaching}. 
CoCoMIC~\cite{ding2023cocomic} and RepoHyper~\cite{phan2024repohyper} enhance code completion capabilities through dependency analysis and learning-based methods but they encounter the problem of difficulty in obtaining training data and poor generalizability. 
% Repoformer~\cite{wu2024repoformer} designs a self-supervised learning approach to enable selective retrieval in code-infilling tasks. 
CodePlan~\cite{bairi2023codeplan}, RepoFuse~\cite{liang2024repofuse} and $A^3\text{-CodeGen}$~\cite{liao2024a3codgen} employ static code analysis to obtain relevant candidates. RepoCoder~\cite{zhang2023repocoder} and De-Hallucinator~\cite{eghbali2024dehallucinator} adopt an approach through iterative retrieval and generation. CodeAgent~\cite{zhang2024codeagent} and ToolGen~\cite{wang2024teaching} explore tool invocation to help code completion. 

Although these efforts show promising performance, \model differs from them in that it does not require labeled data to train and uses a novel stop signal mechanism to know when to retrieve and which candidates to retain. 

\vspace{-2pt}
\section{Conclusion}
\vspace{-2pt}
In this paper, we propose \model, a novel reinforcement learning framework for repository-level code completion. We enable the retriever to iteratively learn by obtaining feedback from the evaluator. Besides, unlike using fixed window candidates or candidates parsing from dependency, we introduce a simple yet effective Split-Aggregate candidate construction method based on human programming habits. Moreover, we propose the stop signal to avoid using useless cross-file context. %, thus leveraging more in-file context. 
% This approach is particularly effective when the in-file context is not sufficient. 
Experimental results indicate that \model achieves state-of-the-art performance on repository-level code completion and demonstrates good generalizability and applicability to further enhancing existing methods.

\section*{Acknowledgements}

The work described in this paper is supported by CCF-Huawei Populus Grove Fund CCF-HuaweiSE202301.